\documentclass[nofootinbib,twocolumn,showpacs,showkeys,preprintnumbers,amsmath,amssymb,floatfix]{revtex4}

\usepackage{graphicx} 
\usepackage{dcolumn} 
\usepackage{bm} 
 
\def\qqbar{\langle \overline{q}q\rangle} 
\def\ssbar{\langle \overline{s}s\rangle}

\def\pslash{p\!\!\!\slash }

\begin{document}

\preprint{}
\title{QCD Sum Rules study of meson-baryon sigma terms}
\author{G\"uray Erkol}
\email{erkol@th.phys.titech.ac.jp}
\author{Makoto Oka}%
\email{oka@th.phys.titech.ac.jp}
\affiliation{Department of Physics, H27, Tokyo Institute of Technology, Meguro, Tokyo 152-8551 Japan}
\author{G\"ursevil Turan}
\email{gsevgur@metu.edu.tr}
\affiliation{Department of Physics, Middle East Technical University, Ankara 06531 Turkey}
\begin{abstract}
	The pion--baryon sigma terms and the strange-quark condensates of the octet and the decuplet baryons are calculated by employing the method of quantum chromodynamics (QCD) sum rules. We evaluate the vacuum-to-vacuum transition matrix elements of two baryon interpolating fields in an external isoscalar-scalar field and use a Monte Carlo--based approach to systematically analyze the sum rules and the uncertainties in the results. We extract the ratios of the sigma terms, which have rather high accuracy and minimal dependence on QCD parameters. We discuss the sources of uncertainties and comment on possible strangeness content of the nucleon and the Delta. 
\end{abstract}

\maketitle
\section{Introduction}

The meson-baryon sigma terms are important for hadron physics as they provide a measure of chiral-symmetry breaking and the scalar quark condensate inside the baryon. In particular, the pion-nucleon and the pion-Delta sigma terms have received much attention and have been extensively analyzed in many problems (see Ref.~\cite{Erkol:2007sj} and references therein). The sigma terms are related to the chiral-symmetry breaking part of the QCD Lagrangian
\begin{equation}\label{lagsb}
	{\cal L}_m=-\hat{m}(\overline{u}u+\overline{d}d)-m_s \overline{s}s=c_1 u_1+c_8 u_8,
\end{equation}
which is expressed in terms of SU(3)-flavor [SU(3)$_F$] singlet and octet pieces 
\begin{align}
	\begin{split}
		&u_1=\overline{u}u+\overline{d}d+\overline{s}s,\\
		&u_8=\overline{u}u+\overline{d}d-2\overline{s}s,
	\end{split}
\end{align}
with $c_1=-(2\hat{m}+m_s)/3$ and $c_8=(m_s-\hat{m})/3$ [$\hat{m}=(m_u+m_d)/2$ is the average light-quark mass]. The strength of the SU(3)$_F$ breaking is controlled by the matrix elements of the octet piece and can be related to the resulting baryon-mass splittings through Gell-Mann--Okubo mass formula. The sigma terms, which are defined in terms of these matrix elements, can be in turn deduced from the SU(3)$_F$ pattern and the observed baryon-mass differences~\cite{Cheng:1975wm, Maiani:1987by, Cheng:1988cz}. 

The sigma terms are equivalent to the values of the scalar form factors
{\allowdisplaybreaks
\begin{subequations}\label{sff}
	\begin{align}
			&\hat{m}\langle {\cal B}(p^\prime)\lvert\overline{u}u+\overline{d}d\rvert {\cal B}(p) \rangle=\sigma_{\cal B}(k) \overline{\upsilon}_{\cal B}(p^\prime)\upsilon_{\cal B}(p),\\[1ex]
			&\hat{m}\langle {\cal B}^\ast(p^\prime,t^\prime) \lvert\overline{u}u+\overline{d}d\rvert {\cal B}^\ast(p,t) \rangle=\label{delmin1}\\
			&-\overline{\upsilon}_{{\cal B}^\ast}^\mu(p^\prime,t^\prime)[g^{\mu\nu} \sigma_{{\cal B}^\ast}(k) +p^{\prime\nu}p^\mu F_T(k)]\upsilon_{{\cal B}^\ast}^\nu(p,t),\nonumber\\[1ex]
			&m_s\langle {\cal B}(p^\prime)\lvert\overline{s}s\rvert {\cal B}(p) \rangle=\sigma^s_{\cal B}(k) \overline{\upsilon}_{\cal B}(p^\prime)\upsilon_{\cal B}(p),\\[1ex]
			&m_s\langle {\cal B}^\ast(p^\prime,t^\prime) \lvert\overline{s}s\rvert {\cal B}^\ast(p,t) \rangle=\label{delmin2}\\
			&-\overline{\upsilon}_{{\cal B}^\ast}^\mu (p^\prime,t^\prime)[g^{\mu\nu} \sigma^s_{{\cal B}^\ast}(k) +p^{\prime\nu}p^\mu F^s_T(k)]\upsilon_{{\cal B}^\ast}^\nu(p,t),\nonumber
	\end{align}%
	\end{subequations}%
}%
at zero momentum transfer, with ${\cal B}=N$, $\Sigma$, $\Lambda$, $\Xi$ and ${\cal B}^\ast=\Delta$, $\Sigma^\ast$, $\Xi^\ast$, $\Omega$. Here, $\upsilon(p,t)$ is the Dirac spinor for the spin-1/2 baryon, $\upsilon^\mu(p,t)$ is the Rarita-Schwinger spin-vector of the spin-3/2 baryon, with the spin projection $t$, $k=(p^\prime-p)^2$ is the momentum transfer, and $\sigma_{{\cal B}^{(\ast)}}(k)$, $\sigma^s_{{\cal B}^{(\ast)}}(k)$ and $F^{(s)}_T(k)$ are the scalar and tensor form factors, respectively. The minus sign on the right-hand sides (RHS) of \eqref{delmin1} and \eqref{delmin2} is conventional like in the case of the free Delta Lagrangian. The sigma terms are also defined via the Feynman-Hellmann theorem as 
{\allowdisplaybreaks
	\begin{subequations}\label{FHt}
	\begin{align} 
		&\sigma_{\cal B}\equiv\sum_{q=u,d}\hat{m} \frac{d\,m_{\cal B}}{d\,m_q}=\hat{m}\langle {\cal B}\lvert\overline{u}u+\overline{d}d\rvert {\cal B} \rangle,\\
		&\sigma_{{\cal B}^\ast}\equiv\sum_{q=u,d}\hat{m}\frac{d\,m_{{\cal B}^\ast}}{d\,m_q} =-\hat{m}\langle {\cal B}^\ast(t) \lvert\overline{u}u+\overline{d}d\rvert {\cal B}^\ast(t^\prime) \rangle, \\
		&\sigma^s_{\cal B}\equiv m_s \frac{d\,m_{\cal B}}{d\,m_s}=m_s\langle {\cal B}\lvert\overline{s}s\rvert {\cal B} \rangle,\\
		&\sigma^s_{{\cal B}^\ast}\equiv m_s\frac{d\,m_{{\cal B}^\ast}}{d\,m_s} =-m_s\langle {\cal B}^\ast(t) \lvert\overline{s}s\rvert {\cal B}^\ast(t^\prime) \rangle,
	\end{align}
	\end{subequations}
}%
where $m_{\cal B}$ and $m_{{\cal B}^\ast}$ denote the octet- and the decuplet-baryon masses, respectively.

The matrix elements $\langle {\cal B}^{(\ast)} \lvert\overline{s}s\rvert {\cal B}^{(\ast)} \rangle$ represent the strangeness content of the baryons, which can be combined with the pion-nucleon and the pion-Delta sigma terms in order to obtain, {\it e.g.}, the eta-baryon sigma terms
\begin{align}\label{sffs}
\begin{split}
	\sigma_{\eta {\cal B}^{(\ast)}}&= \frac{1}{3}\langle {\cal B}^{(\ast)} \lvert \hat{m}(\bar{u}u+\bar{d}d) + 2 m_s \bar{s}s \rvert {\cal B}^{(\ast)} \rangle\\
	&\equiv \frac{1}{3} (\sigma_{{\cal B}^{(\ast)}} +  2 \sigma^s_{{\cal B}^{(\ast)}}).
\end{split}
\end{align}%
The strange quark condensate of the nucleon is of special interest, which is expressed by the ratio
\begin{equation}\label{stfr}
	y=\frac{2\langle N \lvert\overline{s}s\rvert N \rangle}{\langle N \lvert\overline{u}u+\overline{d}d\rvert N \rangle},
\end{equation} 
and related to the pion-nucleon sigma term through
\begin{equation}
	\sigma_N=\sigma^{(0)}_N / (1-y),
\end{equation}
where $\sigma^{(0)}_N=32$~MeV is the pion-nucleon sigma term obtained from the matrix elements of the octet piece in Eq.~\eqref{lagsb} using baryon-mass splittings~\cite{Cheng:1988cz}. Hence, a discrepancy between $\sigma^{(0)}_N$ and the directly observed $\sigma_N$ gives a measure of the strangeness content of the nucleon. The assumption from the Okubo-Zweig-Iizuka (OZI) rule as $y=0$ implies a $\sigma_N$ which is significantly smaller than expectations based on $\pi$-$N$ scattering. The resulting puzzle can be solved by considering the possibility of a non-vanishing $\overline{s}s$ content in the nucleon. This interesting issue has been tackled using various theoretical approaches, {\it e.g.}, the chiral perturbation theory gives $y\simeq 0.21$~\cite{Borasoy:1996bx}, and lattice QCD gives $y \simeq 0.36$~\cite{Dong:1995ec}.

To our knowledge, there are only a few calculations in the literature for the meson-hyperon sigma terms. The quark condensates of the baryons have been calculated in Ref.~\cite{Hatsuda:1994pi} by means of a Nambu--Jona-Lasinio (NJL) approach to QCD. A chiral model has been used in Ref.~\cite{Barros:2002mt} to evaluate the long-range part of the hyperon scalar form factors and the pion--octet-baryon sigma terms. When considered in the framework of the octet and the decuplet baryons, a determination of the the sigma terms is important for understanding the role played by the chiral-symmetry breaking in the octet-decuplet mass splittings. Moreover, since there is no direct coupling of the pion to Lambda baryon, the $\pi$-$\Lambda$ sigma term cannot be directly determined from experiment. Therefore a theoretical determination of the $\pi$-$\Lambda$ sigma term together with the $\pi$-$\Sigma$ sigma term is crucial as these terms are related to $\Lambda$-$\Sigma$ mass splitting. Two of us have recently calculated~\cite{Erkol:2007sj} the pion-nucleon and the pion-Delta sigma terms by utilizing the external-field QCD sum rules (QCDSR), which are a powerful tool to extract qualitative and quantitative information about hadron properties~\cite{Shifman:1978bx,Shifman:1978by,Reinders:1984sr,Ioffe:1983ju}. In this framework, one starts with a correlation function that is constructed in terms of hadron interpolating fields. On the theoretical side, the correlation function is calculated using the Operator Product Expansion (OPE) in the Euclidian region. This correlation function is matched with an {\em Ansatz} that is introduced in terms of hadronic degrees of freedom on the phenomenological side. The matching provides a determination of hadronic parameters like baryon masses, magnetic moments, coupling constants of hadrons, and so on. Our aim in this work is to calculate the scalar quark condensates of the octet and the decuplet baryons and the related sigma terms defined in Eq.~(\ref{sff}), by using the external-field QCD sum rules. To determine the value of the sigma terms, we evaluate the vacuum-to-vacuum transition matrix elements of two baryon interpolating fields in an external isoscalar-scalar field. For our numerical procedure, we use the Monte Carlo--based analysis introduced in Ref.~\cite{Leinweber:1995fn}. This method provides a more systematic treatment of uncertainties in QCDSR.

Our paper is organized as follows: In Section~\ref{secNSR}, we present the
formulation of QCDSR and construct the relevant sum rules. We give the numerical analysis of the sum rules in Section~\ref{secAN}. Finally, we discuss the results and arrive at our conclusions in Section~\ref{secCONC}.

\section{The derivation of the sum rules}~\label{secNSR}
In the external-field QCDSR method, one starts with the correlation function of the baryon interpolating fields in the presence of an external constant isoscalar-scalar field $S_q$, defined by 
	\begin{subequations}\label{cor1a} 
	\begin{align}	
		&i\int d^4 x~ e^{i p\cdot x}\, \left \langle 0\left \lvert{\cal 
		T}[\eta_{\cal B}(x)\overline{\eta}_{\cal B}(0)]\right\rvert 0\right\rangle_{S_q}=\nonumber\\
		&\quad\mathring{\Pi}_{\cal B}(p) + S_q\, \hat{\Pi}^q_{\cal B} (p)+ O(S_q^2),\\ 
		&i\int d^4 x~ e^{i p\cdot x}\, \left \langle 0\left \lvert{\cal T}[\eta_{\cal B^\ast}^\mu(x)\overline{\eta}_{\cal B^\ast}^\nu(0)]\right\rvert 0\right\rangle_{S_q} =\nonumber\\
		&\quad [\mathring{\Pi}_{\cal B^\ast}]^{\mu\nu}(p)+ S_q\, [\hat{\Pi}^q_{\cal B^\ast}]^{\mu\nu} (p)+ O(S_q^2),
	\end{align}	
	\end{subequations}%
where $\eta_{\cal B}$ and $\eta^\mu_{\cal B^\ast}$ are the octet- and the decuplet-baryon interpolating fields, which are respectively given as
\begin{subequations}
	\begin{align} 	
		\begin{split}\label{intfi}
		&\eta_N=\epsilon_{abc}\left[u_a^T C\gamma_\mu u_b\right]\gamma_5 \gamma^\mu d_c,\quad \eta_\Sigma= \eta_N(d\rightarrow s),\\
		&\eta_\Xi= \eta_N(u\rightarrow s), \\
		&\eta_{\Lambda} = \sqrt{\frac{2}{3}}\,\epsilon_{abc}\left\{\left[u_a^T C\gamma_\mu s_b\right]\gamma_5 \gamma^\mu d_c - \left[d_a^T C\gamma_\mu s_b\right] \gamma_5 \gamma^\mu u_c \right \},
		\end{split}\\
		\begin{split}
		&\eta^\mu_\Delta = \epsilon_{abc}\left[u_a^T C\gamma^\mu u_b\right] u_c,\\
		& \eta^\mu_{\Sigma^\ast} = \sqrt{\frac{1}{3}}\,\epsilon_{abc}\left\{2\left[u_a^T C\gamma^\mu s_b\right] u_c + \left[u_a^T C\gamma^\mu u_b\right] s_c \right \}, \\
		&\eta^\mu_\Omega = \eta^\mu_\Delta(u\rightarrow s),\quad \eta^\mu_{\Xi^\ast} = \eta^\mu_{\Sigma^\ast}(s\rightarrow d,\, u\rightarrow s).
		\end{split}
		\end{align}%
\end{subequations}%
Here $a,b,c$ are the color indices, $T$ denotes transposition and $C=i\gamma^2\gamma^0$. For the interpolating fields of the octet baryons, there are two independent local operators, but the ones in Eq.~(\ref{intfi}) are the optimum choices for the lowest-lying positive-parity baryons (see, {\em e.g.}, Ref~\cite{Jido:1996ia} for a discussion on negative-parity baryons in QCDSR). $\mathring{\Pi}_{\cal B}(p)$ and $\mathring{\Pi}_{\cal{B^\ast}}(p)$ are the correlation functions when the external field is absent and correspond to the functions that are used to determine the baryon masses. The second terms in Eqs.~\eqref{cor1a} represent the linear responses of the correlators to a small external scalar field $S_q$, which are computed with an additional term to the QCD Lagrangian:
	\begin{equation}\label{addLag} 
		\Delta{\cal L} = -S\,g^S\left[\overline{u}(x)\,u(x)\,+ \,\overline{d}(x)\,d(x)\right]-S_s\,g_s^S\left[\overline{s}(x)\,s(x)\right].
	\end{equation} 
Here, $S_u= S_d \equiv S$ ($S_s$) represent the external scalar field and $g^S$ ($g_s^S$) is associated with the coupling of the external scalar field to the $u$- and the $d$- ($s$-) quark. $\hat{\Pi}^u_{\cal B^{(\ast)}} =\hat{\Pi}^d_{\cal B^{(\ast)}} \equiv \hat{\Pi}_{\cal B^{(\ast)}}$ ($\hat{\Pi}^s_{\cal B^{(\ast)}}$) denote the correlation functions in the existence of the external $S$ ($S_s$) field. The external scalar field contributes to the correlation functions in Eq.~\eqref{cor1a} in two ways: first, it directly couples to the quark field in the baryon currents and second, it modifies the condensates by polarizing the QCD vacuum. In the presence of an external scalar field there are no correlators that break the Lorentz invariance, like $\langle\overline{q}\sigma_{\mu\nu}q\rangle$ which appears in the case of an external electromagnetic field $F^{\mu\nu}$. However, the correlators already existing in the vacuum are modified by the external field, {\em viz.}
{\allowdisplaybreaks
	\begin{subequations}\label{vaccon} 
		\begin{align}
		\begin{split}	
		&\qqbar_S {}\equiv \qqbar - \chi S \qqbar ,\quad \ssbar_{S} {}\equiv \ssbar - \tilde{\chi} S \ssbar ,\\
		&\qqbar_{S_s} {}\equiv \qqbar - \tilde{\chi}^s S_s \qqbar ,\quad \ssbar_{S_s} {}\equiv \ssbar - \chi^s S_s \ssbar,
		\end{split}\\
		\begin{split}
		&\langle g_c \overline{q} {\bm\sigma}\cdot {\bm G} q\rangle_S {}\equiv \langle g_c \overline{q} {\bm\sigma}\cdot {\bm G} q\rangle - \chi_G S \langle g_c \overline{q} {\bm\sigma}\cdot {\bm G} q\rangle ,\\
		&\langle g_c \overline{s} {\bm\sigma}\cdot {\bm G} s\rangle_{S} {}\equiv \langle g_c \overline{s} {\bm\sigma}\cdot {\bm G} s\rangle - \tilde{\chi}_G S \langle g_c \overline{s} {\bm\sigma}\cdot {\bm G} s\rangle,\\
		&\langle g_c \overline{q} {\bm\sigma}\cdot {\bm G} q\rangle_{S_s} {}\equiv \langle g_c \overline{q} {\bm\sigma}\cdot {\bm G} q\rangle - \tilde{\chi}^s_G S_s \langle g_c \overline{q} {\bm\sigma}\cdot {\bm G} q\rangle ,\\
		&\langle g_c \overline{s} {\bm\sigma}\cdot {\bm G} s\rangle_{S_s} {}\equiv \langle g_c \overline{s} {\bm\sigma}\cdot {\bm G} s\rangle - \chi^s_G S_s \langle g_c \overline{s} {\bm\sigma}\cdot {\bm G} s\rangle ,
		\end{split}
	\end{align}
	\end{subequations}%
}%
where $\chi$ ($\equiv \chi^u \equiv \chi^d$), $\tilde{\chi}$ ($\equiv \tilde{\chi}^u \equiv \tilde{\chi}^d$), $\chi^s$ and $\tilde{\chi}^s$ are the susceptibilities corresponding to the quark condensates. Similarly, $\chi_G$ ($\equiv \chi_G^u \equiv \chi_G^d$), $\tilde{\chi}_G$ ($\equiv \tilde{\chi}_G^u \equiv \tilde{\chi}_G^d$), $\chi_G^s$ and $\tilde{\chi}_G^s$ denote the susceptibilities corresponding to the quark-gluon mixed condensates. Here we explicitly assume that the $u$- and the $d$- ($s$-) quark fields couple solely to the external field $S$ ($S_s$). The quark condensates get modified in the presence of the external fields $S$ and $S_s$ as follows:
	\begin{subequations}\label{chidef}
		\begin{align}
		&\frac{\partial \langle \overline{q}_i q_i \rangle}{\partial m_j}=\chi^i \langle \overline{q}_i q_i \rangle, \quad \frac{\partial \langle g_c \overline{q}_i {\bm\sigma}\cdot {\bm G} q_i \rangle}{\partial m_j}=\chi_G^i \langle g_c \overline{q}_i {\bm\sigma}\cdot {\bm G} q_i \rangle, \nonumber\\ 
		&\quad\text{for}\quad i=j,\\
		&\frac{\partial \langle \overline{q}_i q_i \rangle}{\partial m_j}=\tilde{\chi}^j \langle \overline{q}_i q_i \rangle, \quad \frac{\partial \langle g_c \overline{q}_i {\bm\sigma}\cdot {\bm G} q_i \rangle}{\partial m_j}=\tilde{\chi}_G^j \langle g_c \overline{q}_i {\bm\sigma}\cdot {\bm G} q_i \rangle, \nonumber\\
		&\quad \text{for}\quad i\neq j,
		\end{align}
	\end{subequations}%
where we retain the non-diagonal responses of $\qqbar$ ($\ssbar$) to external $S_s$ ($S_q$) field via the susceptibilities $\tilde{\chi}$ and $\tilde{\chi}^s$, and similarly for the quark-gluon mixed condensates via $\tilde{\chi}_G$ and $\tilde{\chi}_G^s$. The coupling of the external scalar field to the quark is simply taken as $g^S \equiv g_s^S=1$.

At the quark level, we have 
{\allowdisplaybreaks
\begin{widetext}
		\begin{subequations}\label{cor2}
			\begin{align}
		\begin{split}	
			&\left \langle 0\Big \lvert{\cal T} [\eta_N(x)\overline{\eta}_N(0)] \Big \rvert0\right \rangle_{S_q}= 2 i \epsilon^{abc}\epsilon^{a^\prime b^\prime c^\prime} \text{Tr} \left \{{\cal S}_u^{b b^\prime}(x) \gamma_\nu C [{\cal S}_u^{a a^\prime}(x)]^T C \gamma_{\mu}\right \}\gamma_5\gamma^\mu {\cal S}_d^{c c^\prime}(x) \gamma^\nu\gamma_5,
		\end{split}\\[1ex]
		\begin{split}
			&\left \langle 0\Big \lvert{\cal T} [\eta_\Sigma(x)\overline{\eta}_\Sigma(0)] \Big \rvert0\right \rangle_{S_q}= \left \langle 0\Big \lvert{\cal T} [\eta_N(x)\overline{\eta}_N(0)] \Big \rvert0\right \rangle_{S_q} \Big({\cal S}_d\rightarrow {\cal S}_s\Big),
		\end{split}\\[1ex] 
		\begin{split}	
			&\left \langle 0\Big \lvert{\cal T} [\eta_\Xi(x)\overline{\eta}_\Xi(0)] \Big \rvert0\right \rangle_{S_q}= \left \langle 0\Big \lvert{\cal T} [\eta_N(x)\overline{\eta}_N(0)] \Big \rvert0\right \rangle_{S_q} \Big({\cal S}_u\rightarrow {\cal S}_s\Big),
		\end{split}\\[1ex]
		\begin{split}
		\raisetag{20pt}
			&\left \langle 0\Big \lvert{\cal T} [\eta_\Lambda(x)\overline{\eta}_\Lambda(0)] \Big \rvert0\right \rangle_{S_q}=\frac{2}{3}i \epsilon^{abc}\epsilon^{a^\prime b^\prime c^\prime} \left(\text{Tr} \left\{{\cal S}_u^{b b^\prime}(x) \gamma_\nu C [{\cal S}_s^{a a^\prime}(x)]^T C \gamma_{\mu}\right \}\gamma_5\gamma^\mu {\cal S}_d^{c c^\prime}(x) \gamma^\nu\gamma_5 \right.\\
			&\quad+\text{Tr} \left\{{\cal S}_d^{c c^\prime}(x) \gamma_\nu C [{\cal S}_s^{a a^\prime}(x)]^T C \gamma_{\mu}\right \}\gamma_5\gamma^\mu {\cal S}_u^{b b^\prime}(x) \gamma^\nu\gamma_5-\gamma_5 \gamma_\mu {\cal S}_d^{c c^\prime}(x) \gamma_\nu C [{\cal S}_s^{b b^\prime}(x)]^T C\\
			&\left.\quad\times \gamma^\mu {\cal S}_u^{a a^\prime}(x) \gamma^\nu \gamma_5-\gamma_5 \gamma_\mu {\cal S}_u^{a a^\prime}(x) \gamma_\nu C [{\cal S}_s^{b b^\prime}(x)]^T C \gamma^\mu {\cal S}_d^{c c^\prime}(x) \gamma^\nu \gamma_5 \right ),
		\end{split}
	\end{align}
		\end{subequations}%
\end{widetext}%
}%
respectively for $N$, $\Sigma$, $\Xi$ and $\Lambda$, and
{\allowdisplaybreaks
\begin{widetext}
	\begin{subequations}		
	\begin{align}\label{cor2dec}
		\begin{split}
			&\left \langle 0\Big \lvert{\cal T} [\eta_\Delta^\mu(x)\bar{\eta}_\Delta^\nu(0)] \Big \rvert0\right \rangle_{S_q}={}-2 i \epsilon^{abc}\epsilon^{a^\prime b^\prime c^\prime} \Big(\text{Tr} \{S_u^{b b^\prime}(x) \gamma^\nu C [S_u^{a a^\prime}(x)]^T C \gamma^{\mu}\}S_u^{c c^\prime}(x)\\
			&\quad +2 S_u^{b b^\prime}(x) \gamma^\nu C [S_u^{a a^\prime}(x)]^T C \gamma^{\mu} S_u^{c c^\prime}(x)\Big), 
		\end{split}\\[1ex]
		\begin{split}
		\raisetag{20pt}
			&\left \langle 0\Big \lvert{\cal T} [\eta_{\Sigma^\ast}^\mu(x)\overline{\eta}_{\Sigma^\ast}^\nu(0)] \Big \rvert 0\right \rangle_{S_q}= -\frac{2}{3} i \epsilon^{abc}\epsilon^{a^\prime b^\prime c^\prime} \left(\text{Tr} \left\{{\cal S}_s^{b b^\prime}(x) \gamma^\nu C [{\cal S}_u^{a a^\prime}(x)]^T C \gamma^{\mu}\right \}{\cal S}_u^{c c^\prime}(x)\right.\\[1ex]
			&\quad +\text{Tr} \left\{{\cal S}_u^{a a^\prime}(x) \gamma^\nu C [{\cal S}_s^{b b^\prime}(x)]^T C \gamma^{\mu} \right \}{\cal S}_u^{c c^\prime}(x)+\text{Tr} \left \{{\cal S}_u^{a a^\prime}(x) \gamma^\nu C [{\cal S}_u^{c c^\prime}(x)]^T C \gamma^{\mu} \right \}{\cal S}_s^{b b^\prime}(x)\\
			&\quad +2 {\cal S}_u^{c c^\prime}(x) \gamma^\nu C [{\cal S}_u^{a a^\prime}(x)]^T C \gamma^{\mu} {\cal S}_s^{b b^\prime}(x)+2 {\cal S}_u^{a a^\prime}(x) \gamma^\nu C [{\cal S}_s^{b b^\prime}(x)]^T C \gamma^{\mu} {\cal S}_u^{c c^\prime}(x)\\[1ex]
			&\left. \quad +2 {\cal S}_s^{b b^\prime}(x) \gamma^\nu C [{\cal S}_u^{a a^\prime}(x)]^T C \gamma^{\mu} {\cal S}_u^{c c^\prime}(x) \right),
		\end{split}\\
		\begin{split}
			&\left \langle 0\Big \lvert{\cal T} [\eta_{\Xi^\ast}^\mu(x)\overline{\eta}_{\Xi^\ast}^\nu(0)] \Big \rvert 0\right \rangle_{S_q}= \left \langle 0\Big \lvert{\cal T} [\eta_{\Sigma^\ast}^\mu(x)\overline{\eta}_{\Sigma^\ast}^\nu(0)] \Big \rvert 0\right \rangle_S \Big({\cal S}_s\rightarrow {\cal S}_d,\, {\cal S}_u\rightarrow {\cal S}_s\Big),
		\end{split}\\[1ex]
		\begin{split}
			&\left \langle 0\Big \lvert{\cal T} [\eta_\Omega(x)\overline{\eta}_\Omega(0)] \Big \rvert0\right \rangle_{S_q}= \left \langle 0\Big \lvert{\cal T} [\eta_\Delta(x)\overline{\eta}_\Delta(0)] \Big \rvert0\right \rangle_{S_q} \Big({\cal S}_u\rightarrow {\cal S}_s\Big),
		\end{split}
	\end{align}%
	\end{subequations}%
\end{widetext}%
}%
respectively for $\Delta$, $\Sigma^\ast$, $\Xi^\ast$ and $\Omega$.

To calculate the Wilson coefficients, we need the quark propagators in the presence of the external scalar field, which are written as
	\begin{equation}
		{\cal S}_q(x)={\cal S}^0_q(x)+{\cal S}^\prime_q(x).
	\end{equation}
The first term on the RHS is the part of the propagator in the absence of the external field, which is given as 
	\begin{align} 
		\begin{split}
			i~[{\cal S}_q^0]^{ab}&\equiv \langle 0|T[q^a(x)\bar{q}^b(0)]|0\rangle_0\\ 
			&=\frac{i~\delta^{ab}}{2\pi^2 x^4}\hat{x}- \frac{i~\lambda_{ab}^n}{32\pi^2} \frac{g_c}{2} G_{\mu\nu}^n \frac{1}{x^2}(\sigma^{\mu\nu}\hat{x}+\hat{x} \sigma^{\mu\nu}) \\
			&\quad-\frac{\delta^{ab}}{12} \langle\overline{q}q\rangle- \frac{\delta^{ab}x^2}{192}\langle g_c \overline{q}{\bm \sigma}\cdot {\bm G} q\rangle-\frac{m_q \delta^{ab}}{4\pi^2 x^2}\\
			&\quad-\frac{m_q}{32\pi^2}\lambda_{ab}^n g_c G_{\mu\nu}^n\sigma^{\mu\nu} \ln(-x^2)-\frac{\delta^{ab}\langle g_c^2 G^2\rangle}{2^9\times 3 \pi^2}\\
			&\quad\times  m_q x^2 \ln(-x^2)+\frac{i~\delta^{ab}m_q}{48}\langle\overline{q}q\rangle \hat{x}+\frac{i~m_q\delta^{ab}x^2}{2^7 \times 3^2}\\
			&\quad\times\langle g_c \overline{q}{\bm\sigma} \cdot {\bm G} q\rangle\hat{x}+O(m_q^2).
		\end{split}
	\end{align}%
The second term appears in the existence of the external field and is given as
	\begin{align} 
		\begin{split}
			i~[{\cal S}_q^\prime]^{ab} &\equiv \langle 0|T[q^a(x)\overline{q}^b(0)]|0\rangle_{S_q}\\
			&= S_q \left[-\frac{ \delta^{ab}}{4\pi^2x^2} -\frac{1}{32\pi^2}\lambda_{ab}^n g_c G_{\mu\nu}^n\sigma^{\mu\nu} \ln(-x^2)\right.\\
			&\quad-\frac{\delta^{ab}\langle g_c^2 G^2\rangle}{2^9\times 3 \pi^2} x^2 \ln(-x^2)+\frac{i~\delta^{ab}}{48} \langle\overline{q}q\rangle \hat{x}\\
			&\quad + \frac{\delta^{ab}\chi^q}{12}\langle\bar{q}q\rangle + \frac{i~\delta^{ab}x^2}{2^7\times 3^2}\langle g_c \overline{q}{\bm\sigma} \cdot {\bm G} q\rangle\hat{x}\\
			&\quad +\frac{\delta^{ab}x^2}{192}\chi^q_{G} \langle g_c \overline{q}{\bm \sigma}\cdot {\bm G} q\rangle +i\frac{m_q}{4 \pi^2 x^2}\hat{x} \\
			&\quad+\frac{m_q}{96}x^2\langle \overline{q}q\rangle -i \frac{m_q}{48}\chi^q \langle \overline{q}q\rangle \hat{x}\\
			&\left.\quad -\frac{i~m_q\delta^{ab}x^2}{2^7 \times 3^2}\chi_G^q \langle g_c \overline{q}{\bm\sigma} \cdot {\bm G} q\rangle\hat{x}  \right]+O(m_q^2,S_q^2),
		\end{split}
	\end{align}%
when the quark and external field have the same flavor. Otherwise, we have
	\begin{align} 
		\begin{split}
			i~[{\cal S}_q^\prime]^{ab} &\equiv \langle 0|T[q^a(x)\overline{q}^b(0)]|0\rangle_{S_{q^\prime}}\\
			&= S_{q^\prime} \left[\frac{\delta^{ab}\tilde{\chi}^{q^\prime}}{12} \langle\bar{q}q\rangle +\frac{\delta^{ab}x^2}{192}\tilde{\chi}^{q^\prime}_{G} \langle g_c \overline{q}{\bm \sigma}\cdot {\bm G} q\rangle\right.\\
			&\quad-i \frac{m_q}{48}\tilde{\chi}^{q^\prime} \langle \overline{q}q\rangle\hat{x}-\frac{i~m_q\delta^{ab}x^2}{2^7 \times 3^2}\\
			&\left.\quad\times\tilde{\chi}_G^{q^\prime} \langle g_c \overline{q}{\bm\sigma} \cdot {\bm G} q\rangle\hat{x} \right]+O(m_{q^\prime}^2,S_{q^\prime}^2).
		\end{split}
	\end{align}%

The analyticity of the correlation function allows us to write the phenomenological side of the sum rules in terms of a double-dispersion relation of the form
\begin{subequations}\label{phenside}
	\begin{align}
		\text{Re}\,\hat{\Pi}^q_{\cal B}(p)&=\frac{1}{\pi^2}\int^\infty_0 \int^\infty_0 
		\frac{\text{Im}\,\hat{\Pi}_{\cal B}(p)}{(s_1-p^2)(s_2-p^2)} \,ds_1\,ds_2,\\
		\text{Re}\,[\hat{\Pi}^q_{\cal B^{\ast}}]^{\mu\nu}(p)&=\frac{1}{\pi^2}\int^\infty_0 \int^\infty_0 
		\frac{\text{Im}\,[\hat{\Pi}^q_{\cal B^{\ast}}]^{\mu\nu}(p)}{(s_1-p^2)(s_2-p^2)} \,ds_1\,ds_2.
	\end{align}
\end{subequations}%
The ground-state hadron contribution is singled out by utilizing the zero-width approximation, where the hadronic contributions from the Breit-Wigner form to the imaginary part of the correlation function is proportional to the $\delta$-function:
\begin{subequations}~\label{satN}
	\begin{align}
	\begin{split}	
		\text{Im}\,\hat{\Pi}_{\cal B}(p)=&\pi^2\delta(s_1-m_{\cal B}^2) \delta(s_2-m_{\cal B}^2) \left \langle 0\lvert\eta_{\cal B} \rvert {\cal B}(p) \right \rangle \\
		&\times\left\langle {\cal B}(p)\lvert S(\overline{u}u+\overline{d}d) \rvert {\cal B}(p)\right\rangle \left\langle {\cal B}(p) \lvert \overline{\eta}_{\cal B} \rvert 0 \right\rangle\\
		&+\pi^2\delta(s_1-m_{\cal B}^2)\delta(s_2-m_{\cal E}^2)\left\langle 0\lvert\eta_{B} \rvert {\cal B}(p) \right \rangle\\
		&\times \left\langle {\cal B}(p)\lvert S(\overline{u}u+\overline{d}d) \rvert {\cal E}(p)\right\rangle \left \langle {\cal E}(p) \lvert \overline{\eta}_{\cal B} \rvert 0 \right \rangle,
	\end{split}	\\
	\begin{split}	
		\text{Im}\,\hat{\Pi}^s_{\cal B}(p)=&\pi^2\delta(s_1-m_B^2) \delta(s_2-m_{\cal B}^2) \left \langle 0\lvert\eta_{\cal B} \rvert {\cal B}(p) \right \rangle\\
		&\times\left\langle {\cal B}(p)\lvert S^s\,\overline{s}s  \rvert {\cal B}(p)\right\rangle \left\langle {\cal B}(p) \lvert \overline{\eta}_{\cal B} \rvert 0 \right\rangle\\
		&+\pi^2\delta(s_1-m_{\cal B}^2)\delta(s_2-m_{{\cal E}}^2)\left\langle 0\lvert\eta_{B} \rvert {\cal B}(p) \right \rangle\\
		&\times \left\langle {\cal B}(p)\lvert S^s\,\overline{s}s \rvert {\cal E}(p)\right\rangle \left \langle {\cal E}(p) \lvert \overline{\eta}_{\cal B} \rvert 0 \right \rangle.
	\end{split}
	\end{align}%
\end{subequations}%
for spin-1/2 baryons. The correlation functions for the spin-3/2 baryons are similarly expressed. In the presence of external field we have transitions to excited baryon states which are denoted by ${\cal E}$.

The matrix elements of the currents $\eta_{\cal B}$ and $\eta^\mu_{\cal B^\ast}$ between the vacuum and the hadron states are defined as 
\begin{subequations}~\label{overlapN}
	\begin{align}	 
		\langle 0 \lvert \eta_{\cal B} \rvert {\cal B}(p,t) \rangle={}& \lambda_{\cal B} \upsilon(p,t),\\
		\langle 0 \lvert \eta^\mu_{\cal B^\ast} \rvert {\cal B^\ast}(p,t) \rangle={}& \lambda_{\cal B^\ast} \upsilon^\mu(p,t), 
	\end{align}%
\end{subequations}%
respectively for the octet and the decuplet baryons, where $\lambda_{\cal B}$ and $\lambda_{\cal B^\ast}$ are the residues. For the spin-3/2 baryons, we make use of the Rarita-Schwinger spin sum, which is 
	\begin{align}~\label{RSss} 
		\begin{split}
		&\sum_t \upsilon^\mu(p,t) \overline{\upsilon}^\nu(p,t)=\\
		&\quad-\left(g^{\mu\nu}-\frac{1}{3}\gamma^\mu \gamma^\nu-\frac{p^\mu \gamma^\nu-p^\nu\gamma^\mu}{3\,m_{\cal B^\ast}}-\frac{2\,p^\mu p^\nu}{3\,m_{\cal B^\ast}^2}\right)(\pslash+m_{\cal B^\ast})\\
		&\quad\equiv T^{\mu\nu}(\pslash+m_{\cal B^\ast}),
		\end{split} 
	\end{align}
where the slash denotes $\pslash=p_\mu\gamma^\mu$. Inserting Eq.~\eqref{overlapN} into Eq.~\eqref{satN} and using the definitions in Eq.~\eqref{sff}, the pole structures of the correlation functions in Eq.~\eqref{phenside} are obtained as
	\begin{align}
		\begin{split}~\label{phen2o} 
			&\lambda_{\cal B}^2 \frac{\pslash+m_{\cal B}}{p^2-m_{\cal B}^2}\,\frac{\sigma^q_{\cal B}}{m_q}\,\frac{\pslash+m_{\cal B}}{p^2-m_{\cal B}^2}\\
			 &\quad+\lambda_{\cal B}\lambda_{\cal E}\frac{\pslash+m_{\cal B}}{p^2-m_{\cal B}^2}\, \alpha_{{\cal B} {\cal E}}\,\frac{\pslash+m_{{\cal E}}}{p^2-m_{{\cal E}}^2},
		\end{split} 
	\end{align}
for spin-1/2 baryons and
	\begin{align}
		\begin{split}~\label{phen2d}
			&-\lambda_{{\cal B}^\ast}^2 \frac{T^{\mu\rho}(\pslash+m_{{\cal B}^\ast})}{p^2-m_{{\cal B}^\ast}^2}\, \frac{\sigma^q_{{\cal B}^\ast}}{m_q}\,g_{\rho\beta}\,\frac{T^{\beta\nu} (\pslash+m_{{\cal B}^\ast})}{p^2-m_{{\cal B}^\ast}^2}\\
			 &\quad+\lambda_{{\cal B}^\ast}\lambda_{{{\cal E}^\ast}} \frac{T^{\mu\rho}(\pslash+m_{{\cal B}^\ast})}{p^2-m_{{\cal B}^\ast}^2}\, \alpha_{{\cal B}{{\cal E}^\ast}}\,g_{\rho\beta}\frac{T^{\beta\nu} (\pslash+m_{{{\cal E}^\ast}})}{p^2-m_{{{\cal E}^\ast}}^2},
		\end{split} 
	\end{align}
for spin-3/2 baryons. Here, the second terms are associated with the transitions to higher baryon states and $\alpha_{{\cal B}{{\cal E}^{(\ast)}}}$ denote the transition matrix elements.

We can bring the correlation functions $\hat{\Pi}_B^q$ and $\hat{\Pi}^q_{B^\ast}$ into the form
\begin{subequations}\label{str} 
	\begin{align}	
		\hat{\Pi}^q_{\cal B}(p)={}&\Pi^q_{\cal B}(p^2)\, \pslash+\Pi_{\cal B}^{q\prime}(p^2),\\
		[\hat{\Pi}_{\cal B^\ast}^{q}]^{\mu\nu}(p)={}& \Pi^q_{\cal B^\ast}(p^2)\,g^{\mu\nu}\,\pslash\,+\, \Pi^{q\prime}_{\cal B^\ast}(p^2)\, g^{\mu\nu} +\dotsb,
	\end{align}%
\end{subequations}%
where the ellipsis represents the Lorentz-Dirac structures other than $g_{\mu\nu}$ and $g_{\mu\nu}\pslash$. Note that one can obtain the sum rules at different Lorentz structures. Here, we choose to work with the sum rules at the structures $\pslash$ and $g^{\mu\nu}\,\pslash$ for the octet and the decuplet baryons, respectively, where the latter are completely contributed by the decuplet baryons with $J=\frac{3}{2}$ (see, {\it e.g.}, Ref.~\cite{Ioffe:1983ju} for details).

One then expresses the correlation function for the octet baryons as a sharp resonance plus a continuum after Borel transformation:
\begin{subequations}\label{phensideN1}
	\begin{align}
		\Pi_{\cal B}(M^2)=&\left(2\lambda_{\cal B}^2 m_{\cal B} \frac{\sigma_{\cal B}}{\hat{m}}+\,C_{\cal B}\,M^2\right)\, \,\frac{e^{-m_{\cal B}^2/M^2}}{M^4}\nonumber\\
		&+\frac{1}{\pi} \int_{w_B^2}^\infty\,\frac{\text{Im} \,\Pi_{\cal B}}{M^4}\, e^{-s_0/M^2}\,ds_0,\\
		\Pi^s_{\cal B}(M^2)=&\left(2\lambda_{\cal B}^2 m_{\cal B} \frac{\sigma^s_{\cal B}}{m_s}+\,C^s_{\cal B}\,M^2\right)\, \,\frac{e^{-m_{\cal B}^2/M^2}}{M^4}\nonumber\\
		&+\frac{1}{\pi} \int_{(w^s_B)^2}^\infty\,\frac{\text{Im} \,\Pi^s_{\cal B}}{M^4}\, e^{-s_0/M^2}\,ds_0.
	\end{align}%
\end{subequations}%
Similarly for decuplet baryons, we write
\begin{subequations}\label{phensideD1}
	\begin{align}
		\Pi_{\cal B^\ast}(M^2)&=-\left(2\lambda_{\cal B^\ast}^2 m_{\cal B^\ast} \frac{\sigma_{\cal B^\ast}}{m_q}+\,C_{\cal B^\ast} \,M^2\right)\,\frac{e^{-m_{\cal B^\ast}^2/M^2}}{M^4}\nonumber\\
		&\quad+\frac{1}{\pi} \int_{w_{B^\ast}^2}^\infty\,\frac{\text{Im}\,\Pi_{\cal B^\ast}}{M^4}\, e^{-s_0/M^2}\,ds_0,\\
		\Pi^s_{\cal B^\ast}(M^2)&=-\left(2\lambda_{\cal B^\ast}^2 m_{\cal B^\ast} \frac{\sigma^s_{\cal B^\ast}}{m_s}+\,C^s_{\cal B^\ast} \,M^2\right)\,\frac{e^{-m_{\cal B^\ast}^2/M^2}}{M^4}\nonumber\\
		&\quad+\frac{1}{\pi} \int_{(w^s_{B^\ast})^2}^\infty\,\frac{\text{Im}\,\Pi^s_{\cal B^\ast}}{M^4}\, e^{-s_0/M^2}\,ds_0.
	\end{align}%
\end{subequations}
Here $w_{B^{(\ast)}}$ and $w^{s}_{B^{(\ast)}}$ denote the continuum thresholds in the existence of the $S_q$ and $S_s$ fields, respectively. Above we have defined $\Pi^u_{\cal B^{(\ast)}} =\Pi^d_{\cal B^{(\ast)}} \equiv \Pi_{\cal B^{(\ast)}}$. We have included the single-pole contributions with the factors $C_{\cal B^{(\ast)}}$ and $C^s_{\cal B^{(\ast)}}$, which correspond to transition strengths to higher baryon states [second terms on the RHS of \eqref{phen2o} and \eqref{phen2d} upon Borel transformation]. These transition terms are not properly suppressed after the Borel transformation and should be included on the phenomenological side.
	
The QCD sum rules are obtained by matching the OPE sides with the hadronic sides and applying the Borel transformation. We give the resulting sum rules in Appendix~\ref{app1}, where we have defined the quark condensate $a_q=-(2\pi)^2\langle\overline{q}q\rangle$, and the quark-gluon--mixed condensate $\langle\overline{q}g_c {\bm \sigma} \cdot {\bf G} q\rangle=m_0^2 \langle\overline{q}q\rangle$ with the QCD coupling-constant squared $g_c^2=4\pi\alpha_s$. The flavor-symmetry breaking is accounted for by the factor $f=\langle\overline{s}s\rangle/\langle\overline{q}q\rangle$ and the four-quark condensate is parameterized as $\langle(\overline{q}q)^2 \rangle\equiv\kappa \langle\overline{q}q\rangle^2$. The continuum contributions are included via the factors 
\begin{subequations}
	\begin{align}
		E^{B^{(\ast)}}_n\equiv 1- \left(1+x+...+ \frac{x^n}{n!}\right) e^{-x},\\ 
		\tilde{E}^{B^{(\ast)}}_n\equiv 1- \left(1+\tilde{x}+...+ \frac{\tilde{x}^n}{n!}\right) e^{-\tilde{x}},
	\end{align}%
\end{subequations}%
with $x=w_{B^{(\ast)}}^2/M^2$ and $\tilde{x}=(w^{s}_{B^{(\ast)}})^2/M^2$. In the sum rules, the third terms on the RHS give the contributions that come from the responses of the continuum thresholds to the external field. Here, $\delta (w_{B^{(\ast)}})^2$ and $\delta (w^s_{B^{(\ast)}})^2$ represent the variations of the continuum threshold and the coefficients are calculated by differentiating the continuum parts of the chiral-even octet and decuplet mass sum rules with respect to the quark mass. These terms are suppressed as compared to the single-pole terms, nevertheless, should be included on the phenomenological side if they are large (see Ref.~\cite{Ioffe:1995jt} for a detailed explanation of this term). The corrections that come from the anomalous dimensions of various operators are included with the factors $L=\log(M^2/\Lambda_{QCD}^2)/\log(\mu^2/\Lambda_{QCD}^2)$, where $\mu=500$~MeV is the renormalization scale and $\Lambda_{QCD}$ is the QCD scale parameter. A variation of the renormalization scale as well as that of the QCD scale parameter has little effect on the results.

\section{Analysis of the sum rules}~\label{secAN}
We determine the uncertainties in the extracted parameters via the Monte Carlo--based analysis introduced in Ref.~\cite{Leinweber:1995fn}. In this analysis, randomly selected, Gaussianly distributed sets are generated from the uncertainties in the QCD input parameters. Here we use $a_q=0.52 \pm 0.05$~GeV$^3$, $b\equiv\left\langle g_c^2 G^2\right\rangle=1.2 \pm 0.6$~GeV$^4$, $m_0^2=0.72 \pm 0.08$~GeV$^2$, and $\Lambda_{QCD}=0.15 \pm 0.04$~GeV. The factorization violation in the four-quark operator is searched via the parameter $\kappa$, where we take $\kappa=2 \pm 1$ and $1 \leq \kappa \leq 4$; here $\langle (\overline{q}q)^2\rangle \geq \langle \overline{q}q\rangle^2$ is assumed via the cut-off at 1. The flavor-symmetry breaking parameter and the mass of the strange quark are taken as $f\equiv \langle \overline{s}s \rangle/\langle \overline{u}u \rangle= 0.83 \pm 0.05$ and $m_s=0.11 \pm 0.02$~GeV, respectively (for a discussion on QCD parameters see, {\it e.g.}, Ref.~\cite{Leinweber:1995fn}). 

The value of the susceptibility $\chi$ can be calculated by using the two-point function~\cite{Jin:1993nn,Erkol:2005jz,Erkol:2006eq}
\begin{align}
	\begin{split}~\label{susc}
   		T(p^2)=&i\int d^4 x e^{i p\cdot x} \langle 0|{\cal T}[\bar{u}(x)u(x)+\bar{d}(x) d(x),\bar{u}(0)u(0)\\
&+\bar{d}(0) d(0)]|0\rangle\, ,
	\end{split}
\end{align}
via the relation
	\begin{equation}\label{suscorr}
		\chi \qqbar=\frac{1}{2} T(0) .
	\end{equation}
The two-point function in Eq.~(\ref{susc}) at $p^2=0$ has been studied in chiral perturbation theory~\cite{Gasser:1983yg} with the result
	\begin{equation}
		\chi=\frac{\qqbar}{16\pi^2 f_\pi^4}\Big(\frac{2}{3}
		\bar{\ell}_1+\frac{7}{3}\bar{\ell}_2-\frac{11}{6}\Big)\, ,
	\end{equation}
where $f_\pi=93$ MeV is the pion decay constant and $\bar{\ell}_1$ and $\bar{\ell}_2$ are the low-energy constants appearing in the effective chiral Lagrangian. A recent analysis of $\pi$-$\pi$ scattering gives $\bar{\ell}_1= -1.9\pm 0.2$ and $\bar{\ell}_2=5.25\pm 0.04$~\cite{Colangelo:2001df}. Using these values of $\bar{\ell}_1$ and $\bar{\ell}_2$ and taking the quark condensate $a_q=0.52\pm 0.05$ GeV$^3$, we find $\chi=-10\pm 1$ GeV$^{-1}$. The susceptibility $\chi_G$ is less certain. It is reasonable to assume $\chi_G\equiv \chi$, however we adopt a larger uncertainty for $\chi_G$ and take $\chi_G=-10 \pm 3 ~\text{GeV}^{-1}$. As we shall see below, the final results are insensitive to a variation in $\chi_G$.

The susceptibilities $\chi^s$ and $\chi^s_G$ can be related to $\chi$ and $\chi_G$, respectively, in a straightforward way by using the three-flavor NJL model~\cite{Kunihiro:1990ts,Hatsuda:1994pi}: In NJL model, the constituent-quark mass is composed of the current-quark mass and a dynamical part ($M^D$) that has a purely non-perturbative origin:  
	\begin{align}\label{consq}
		\begin{split}
		M_i&=m_i+M_i^D; \qquad i=u,\,d,\,s,\\
		M_i^D&=-2\,g_s \langle \overline{q}_i q_i \rangle + M_i^\prime,
	\end{split}
	\end{align}%
where $m_i$ is the current-quark mass, $g_s$ is the four-quark coupling and $M_i^\prime$ represents a potentially small contribution that originates from the six-quark coupling. The scalar charge of the constituent quark can be defined by using the Feynman-Hellmann theorem as
	\begin{equation}
		Q_{ij}=\langle M_i | \overline{q}_j q_j  | M_i \rangle = \frac{\partial M_i}{\partial m_j}=\delta_{ij} + R_{ij},
	\end{equation}
where we have defined $R_{ij}=\partial M_i^D/\partial m_j$ with the aid of Eq.~\eqref{vaccon}. In NJL model, $\partial M_i^\prime/\partial m_i$ vanishes in the diagonal case and we obtain $R_{ii}= -2\,g_s\,\partial \langle \overline{q}_i q_i \rangle/\partial m_i$. Using the values of $R_{ij}$ as given in Eq.~(4.13) of Ref.~\cite{Hatsuda:1994pi}, we find 
	\begin{equation}
		\frac{R_{uu}}{R_{dd}}=1;\qquad \frac{R_{uu}}{R_{ss}}\equiv \frac{\chi}{\chi^s\,f}=\frac{1.14}{0.42}\simeq 2.7,
	\end{equation}
which implies that $\chi^s<\chi$. Inserting the central values as $\chi=-10$~GeV$^{-1}$ and $f=0.83$ yields $\chi^s \simeq -4.5$~GeV$^{-1}$. In our numerical analysis we consider the susceptibility values $\chi^s=-4\pm 1$~GeV$^{-1}$ by allowing a generous uncertainty and adopt $\chi^s_G\equiv\chi^s$. 

The non-vanishing values of the non-diagonal susceptibilities $\tilde{\chi}$, $\tilde{\chi}_G$, $\tilde{\chi}^s$ and $\tilde{\chi}_G^s$, which the sum rules for $\sigma^s_N$, $\sigma^s_\Delta$ and $\sigma_\Omega$ solely depend on, lead to some $\overline{s}s$ content for the nucleon and the Delta, and to $\overline{u}u$ ($\overline{d}d$) content for the Omega. Such anomalous quark contents are OZI-rule suppressed, therefore we expect that these non-diagonal susceptibilities should be very small, if non-zero. $\tilde{\chi}$ and $\tilde{\chi}^s$ can be expressed in terms of a correlation function as in Eq.~\eqref{susc} and estimated in chiral perturbation theory. In our analysis, we shall treat these susceptibilities as free parameters and adopt the ranges $-2 \leq \tilde{\chi} \leq -1$~GeV$^{-1}$ and $-2 \leq \tilde{\chi}^s \leq -1$~GeV$^{-1}$. We also assume $\tilde{\chi} \equiv \tilde{\chi}_G$ and $\tilde{\chi}^s \equiv \tilde{\chi}_G^s$. We would like to note that these ranges are consistent with those in Ref.~\cite{Jin:1995fy}, which were taken to reproduce the baryon isospin mass splittings. Moreover, since the sum rules for $\sigma^s_N$ depend solely on the non-diagonal susceptibilities, $\tilde{\chi}$ and $\tilde{\chi}_G$ are actually constrained by the strangeness content of the nucleon, which can be determined independently by using other approaches. As we shall see below, these values of the susceptibilities produce a strangeness content for the nucleon in agreement with the expectations based on lattice QCD and chiral perturbation theory. 

We use $10^3$ such configurations from which the uncertainty estimates in the extracted parameters are obtained using a fit of the LHS of the sum rules to the RHS. For $\hat{m}$, we make use of the Gell-Mann--Oakes--Renner relation, which is 
	\begin{equation}
		\label{GOR} 2\hat{m}\langle\overline{q}q\rangle = -m_\pi^2 f_\pi^2,
	\end{equation}
where $m_\pi=138~\text{MeV}$ is the pion mass. We use the chiral-odd mass sum rules given in Appendix~\ref{app2} for normalization of the sigma-term sum rules,
which have been found to be more reliable than the chiral-even sum rules~\cite{Leinweber:1995fn,Lee:1997ix}. The Monte Carlo analyses of the sum rules are performed by first fitting the mass sum rules \eqref{Nmass}-\eqref{OmegamassSR} to obtain the pole residues $\tilde{\lambda}_{\cal B}$ and $\tilde{\lambda}_{\cal B^\ast}$, and these residue values are used in the sigma-term sum rules \eqref{sigsigma}-\eqref{sigomega} for each corresponding parameter set. 

The valid Borel regions are determined so that the highest-dimensional operator contributes no more than about $10\%$ to the OPE side, which gives the lower limit on the valid Borel region and ensures OPE convergence. The upper limit is determined using a criterion such that the continuum-plus-continuum-change and plus-excited-state contributions are less than about $50\%$ of the phenomenological side, which is imposed so as to warrant the pole dominance (this constraint is slightly released for $\sigma_\Delta$ sum rule~\cite{Erkol:2007sj}). Note that, while the first criterion is rather straightforward, one does not initially have a complete control on the second, since the phenomenological parameters are determined from the fit and they are correlated. We use the following strategy: we first make the fits in a reasonably selected Borel region, which is then adjusted by trial and error according to the fit results until the above criteria are satisfied.

\begin{table}[t]
	\caption{The parameter values that we use for the numerical analysis of the mass sum rules \eqref{Nmass}-\eqref{OmegamassSR}, the obtained values of the overlap amplitudes and the continuum contributions for each sum rule at the lower ends of the valid Borel regions.}
\setlength{\extrarowheight}{2pt}
\begin{tabular}{cccccc}
		\hline\hline Res. & Region & cont. & $m$~(GeV) & $w_{B^{(\ast)}}$~(GeV) & $\tilde{\lambda}_B^2$~(GeV$^6$)  \\[1ex]
		\hline
		N & [0.9-1.2]& $24\%$ & 0.939 & 1.5 & $1.64 \pm 0.18$  \\
		$\Lambda$ & [0.9-1.4]& $13\%$ & 1.116 & 1.7 & $2.96 \pm 0.30$  \\ 
		$\Sigma$ & [0.8-1.3]& $30\%$ & 1.189 & 1.7 & $2.90 \pm 0.32$  \\
		$\Xi$ & [0.9-1.4]& $44\%$ & 1.321 & 1.7 & $3.76 \pm 0.44$  \\
		$\Delta$ & [1.0-1.2]& $30\%$ & 1.47 & 1.7 & $5.04 \pm 1.04$  \\
		$\Sigma^\ast$ & [0.9-1.3] & $13\%$ & 1.385 & 1.8 & $5.36 \pm 0.88$  \\
		$\Xi^\ast$ & [0.9-1.4] & $7\%$ & 1.533 & 2.0 & $8.38 \pm 1.24$  \\
		$\Omega$ & [1.0-1.6] & $6\%$ & 1.672 & 2.3 & $12.70 \pm 1.80$   \\[1ex]
		\hline\hline
	\end{tabular}
	\label{masses}
\end{table}

\begin{table*}[t]
	\caption{The values of the sigma terms, $\sigma^{(s)}_{B^{(\ast)}}$, and the transition strengths, $\tilde{C}^{(s)}_{B^{(\ast)}}$, for each resonance as obtained from a fit of $10^3$ parameter sets. The second error in $\sigma^{(s)}_{B^{(\ast)}}$ is due to the uncertainty in the $\kappa$ parameter and the first is the sum of errors due to all remaining sources. The second column shows the valid Borel regions and the third column shows the continuum-plus-continuum-change and plus-excited-state contributions at the lower ends of the valid Borel regions. We give the ratios of the sigma terms in the last column.}
\setlength{\extrarowheight}{2pt}
\begin{tabular*}{1.0\textwidth}{@{\extracolsep{\fill}}ccccccccc}
		\hline\hline 
		Resonance & Region & cont.+exc. & $\tilde{C}_{B}$~(GeV$^5$)& \multicolumn{3}{c}{$\sigma_{B}$~(MeV)} & $\sigma_B/\sigma_\Lambda$  \\[1ex]
\cline{5-7}	 &  &  & & QCDSR & Ref.~\cite{Hatsuda:1994pi} & Ref.~\cite{Barros:2002mt} \\[1ex]
		\hline 
		$\Lambda$ & [0.9-1.8]& $14\%$ & $1.32(1.15)$  & $14(03)(02) $ & 45.4 & 33.5 & 1\\
		N & [0.9-1.3]& $30\%$ & $7.45(2.92)$  & $53(09)(15) $ & 56.1 & 46.0 & 3.73(67)\\
		$\Sigma$ & [1.1-1.7]& $30\%$ & $7.91(2.96)$ & $43(07)(08) $ & 33.3 & 29.2 & 3.07(62)\\
		$\Xi$ & [0.8-1.2]& $28\%$ & $-3.50(1.03)$  & $7(2)(2) $ & 27.1 & 12.0 &0.44(16) \\[1ex]
		Resonance & Region & cont.+exc. & $\tilde{C}_{B^\ast}$~(GeV$^5$)& \multicolumn{3}{c}{$\sigma_{B^\ast}$~(MeV)}& $\sigma_{B^\ast}/\sigma_{\Sigma^\ast}$   \\
		$\Sigma^\ast$ & [1.2-1.6] & $36\%$ & $24.25(6.49)$ & $56(10)(12)$ & 23 & -  & 1\\
		$\Delta$ & [1.3-1.5]& $56\%$ & $39.94(12.65)$  & $54(10)(15) $ & 27.6 & - & 0.96(09)\\
		$\Xi^\ast$ & [1.2-1.7] & $25\%$ & $11.80(2.89)$ & $28(04)(06)$ & 17 & - & 0.51(03)\\
		$\Omega$ & [1.3-2.1] & $21\%$ & $2.33(1.41)$ & $6(1)(1)$ & 9 & - & 0.11(02) \\[1ex]
	 Resonance & Region & cont.+exc. & $\tilde{C}^s_{B}$~(GeV$^5$)& \multicolumn{3}{c}{$\sigma^s_{B}$~(MeV)} & $\sigma^s_B/\sigma^s_\Lambda$ \\[0.5ex]
		$\Lambda$ & [0.9-1.4]& $21\%$ & $1.95(52)$  & $243(72)(31) $ & 191& - & 1 \\
		N & [0.9-1.4]& $28\%$ & $0.93(47)$  & $161(41)(25) $ & 58 & - & 0.69(19)\\
		$\Sigma$ & [0.8-1.2]& $30\%$ & $-2.81(39)$ & $129(36)(34) $ & 215 & - & 0.58(25) \\
		$\Xi$ & [1.0-1.4]& $17\%$ & $-2.44(47)$  & $272(99)(44)$ & 341	& -	& 1.18(16)		
\\[1ex]
		Resonance & Region & cont.+exc. & $\tilde{C}^s_{B^\ast}$~(GeV$^5$)& \multicolumn{3}{c}{$\sigma^s_{B^\ast}$~(MeV)} & $\sigma^s_{B^\ast}/\sigma^s_{\Sigma^\ast}$   \\
		$\Sigma^\ast$ & [1.2-1.6] & $35\%$ & $7.24(2.23)$ & $301(080)(056)$ & 138  & - & 1\\
		$\Delta$ & [1.2-1.5] & $41\%$ & $4.25(1.66)$ & $173(053)(034)$ &  29 & - & 0.57(12)\\
		$\Xi^\ast$ & [1.2-1.8] & $30\%$ & $10.38(3.19)$ & $323(091)(048)$ & 276 & - & 1.09(12)\\
		$\Omega$ & [1.1-1.8] & $16\%$ & $11.03(2.91)$ & $380(117)(048)$ & 408 & - & 1.29(22)\\[1ex]
		\hline\hline
	\end{tabular*}
	\label{ST_table}
\end{table*}

The parameter values that we use for the numerical analysis of the mass sum rules \eqref{Nmass}-\eqref{OmegamassSR}, together with the fitted values of the overlap amplitudes, are given in Table~\ref{masses}. We also give the continuum contributions for each sum rule at the lower ends of the valid Borel regions (continuum contributions amount to 50\% of the total phenomenological side at the higher ends of the Borel regions). In order to  reduce the uncertainties in the final results as much as possible, we fix the baryon masses at their experimental values and the continuum thresholds at around the first-excited resonance masses as suggested by the Particle Data Group~\cite{Yao:2006px}. It is well-known that the chiral-odd sum rules are less prone to higher order corrections in $\alpha_s$~\cite{Chung:1984gr} and they perform better as compared to the chiral-even sum rules due to cancellations in the continuum, however the chiral-odd sum rule somewhat overestimates the mass of the Delta resonance (see Refs.~\cite{Erkol:2007sj,Lee:1997ix,Erkol:2008gp} for details). Therefore, we take as input the value of the Delta mass as suggested by the chiral-odd sum rule.
\begin{figure}
	[tbh] 
	\caption{The subtracted form of the sum rules~\eqref{sigsigma}, \eqref{siglambda}, and~\eqref{sigxi}. The solid line is the double-pole contribution and the dashed-line is the OPE-minus-excited states and minus-continuum-change contributions, where we use the average values of the QCD and the obtained fit parameters. The error bars at the two ends represent the uncertainties in the QCD parameters.}
	\includegraphics[scale=0.40]{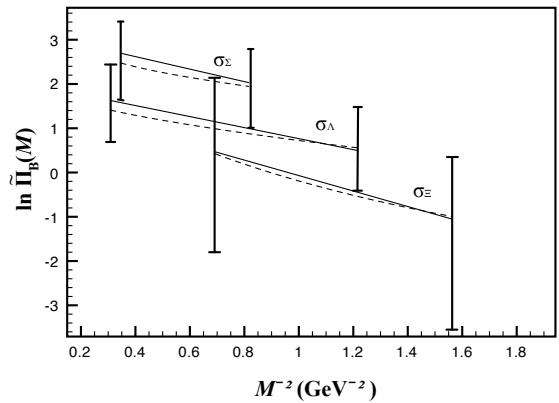} \label{log_octet}
\end{figure}

\begin{figure}
	[tbh] 
	\caption{(color online) Same as Fig.~\ref{log_octet} but for the sum rules~\eqref{sigsigmaast},~\eqref{sigxiast} and~\eqref{sigomegaud}. The error bar at the higher end for $\sigma_{\Xi^\ast}$ is slightly shifted for clear viewing.}
	\includegraphics[scale=0.40]{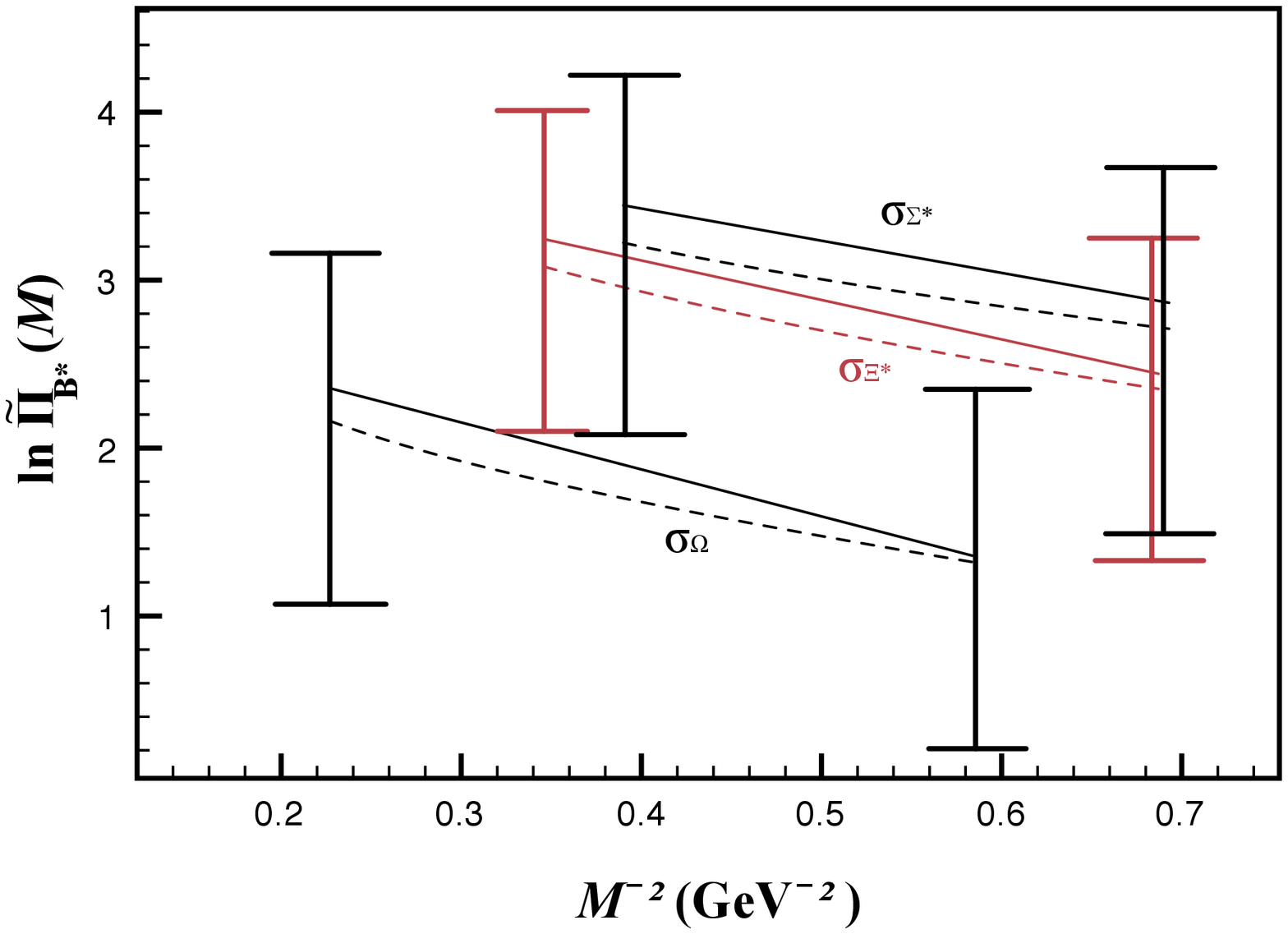}  \label{log_decuplet}
\end{figure}

It is relevant to point out that the dominant contributions to the OPE sides of the sum rules \eqref{sigsigma}-\eqref{sigomega} come from the terms that involve the susceptibilities, whereas the terms that the continuum effects enter with, either do not involve the susceptibility or are proportional to quark masses, which are suppressed as compared to the leading OPE terms. This leads also to a suppression of the continuum contributions. Since the continuum contributions in the sum rules \eqref{sigsigma}-\eqref{sigomega} are suppressed as compared to the total phenomenological side, it becomes difficult to extract information about the continuum thresholds from the fit. Therefore, we have assumed that the continuum thresholds are equivalent to those for the mass sum rules. We also take $w_{B^{(\ast)}}\equiv w_{B^{(\ast)}}^s$. The variation of the continuum thresholds, $\delta (w_{B^{(\ast)}})^2$ and $\delta (w^s_{B^{(\ast)}})^2$, can also be determined from the fit. However, instead of taking these as free parameters, we proceed with a generous assumption that the continuum thresholds change by $25\%$ with the external field \emph{viz.} $\delta (w_{B^{(\ast)}})^2=w_{B^{(\ast)}}/4$ and $\delta (w^s_{B^{(\ast)}})^2=w^s_{B^{(\ast)}}/4$. We observe that such changes minimally contribute to the final results therefore can safely be neglected.
\begin{figure}
	[tbh] 
	\caption{Same as Fig.~\ref{log_octet} but for the sum rules~\eqref{sigN} and~\eqref{sigxis}.}
	\includegraphics[scale=0.40]{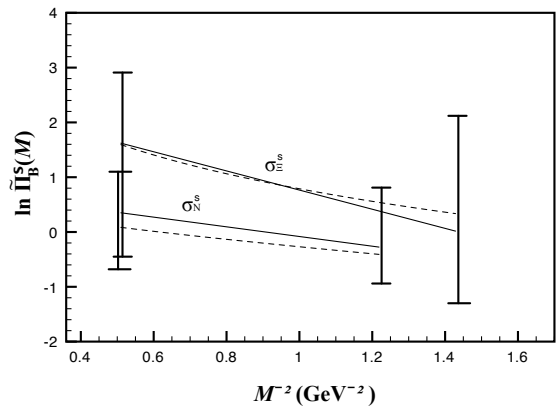}  \label{log_octet_s2}
\end{figure}

\begin{figure}
	[tbh] 
	\caption{Same as Fig.~\ref{log_octet} but for the sum rules~\eqref{sigsigmas} and \eqref{siglambdas}.}
	\includegraphics[scale=0.40]{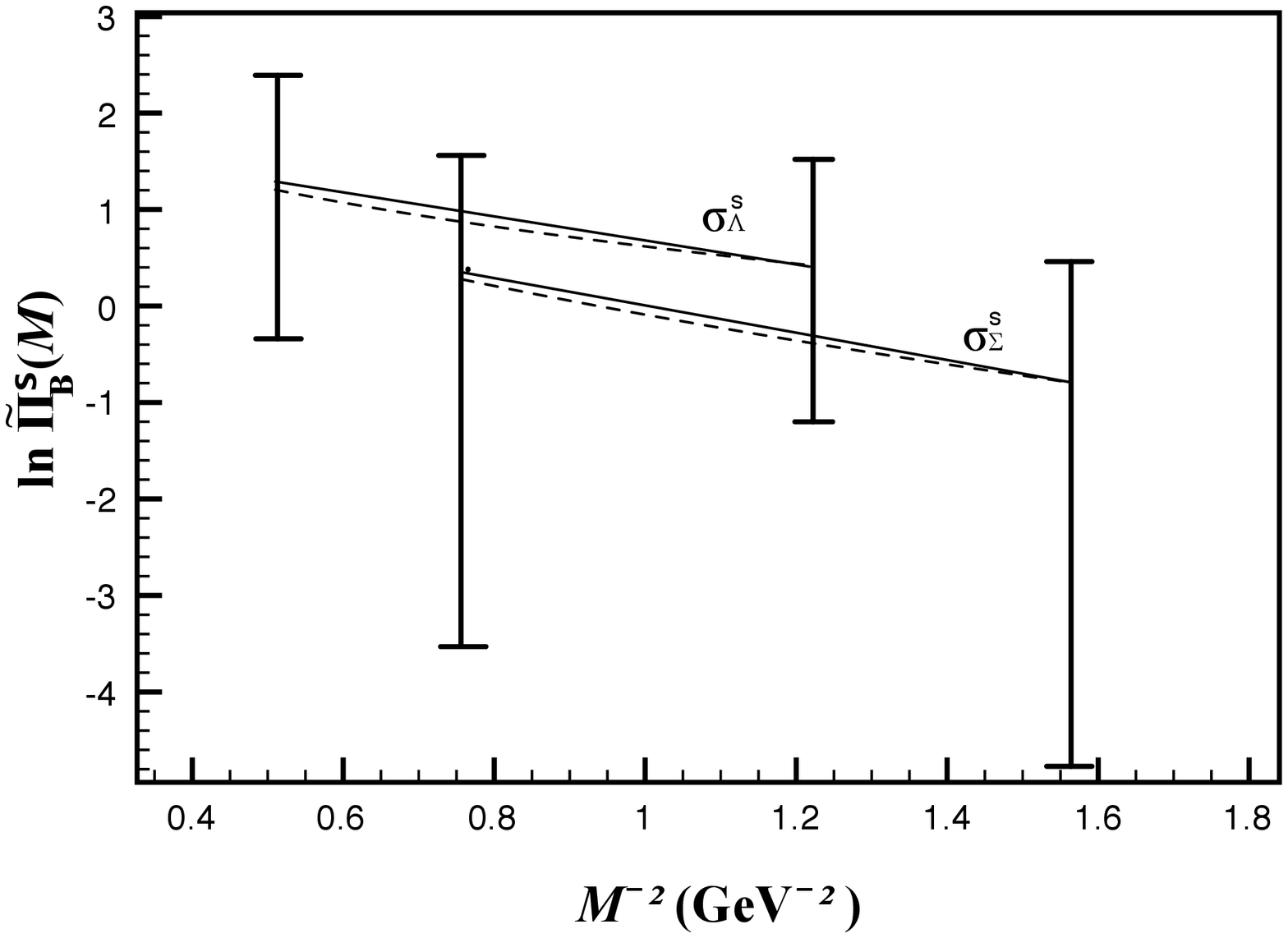}  \label{log_octet_s}
\end{figure}

\begin{figure}
	[tbh] 
	\caption{Same as Fig.~\ref{log_octet} but for the sum rules~\eqref{sigDeltas} and~\eqref{sigxiasts}. The error bar at the higher end for $\sigma^s_{\Delta}$ is slightly shifted for clear viewing.}
	\includegraphics[scale=0.40]{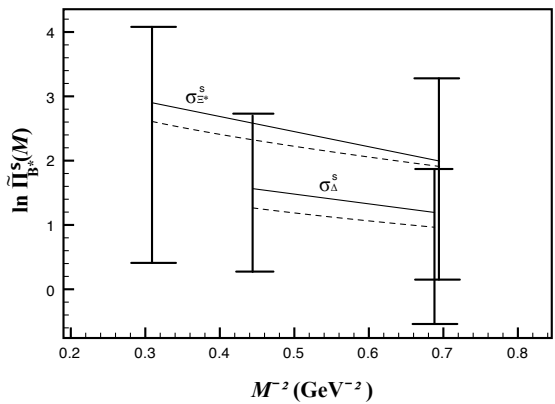}  \label{log_decuplet_s2}
\end{figure}

\begin{figure}
	[tbh] 
	\caption{Same as Fig.~\ref{log_octet} but for the sum rules~\eqref{sigsigmaasts} and~\eqref{sigomega}.}
	\includegraphics[scale=0.40]{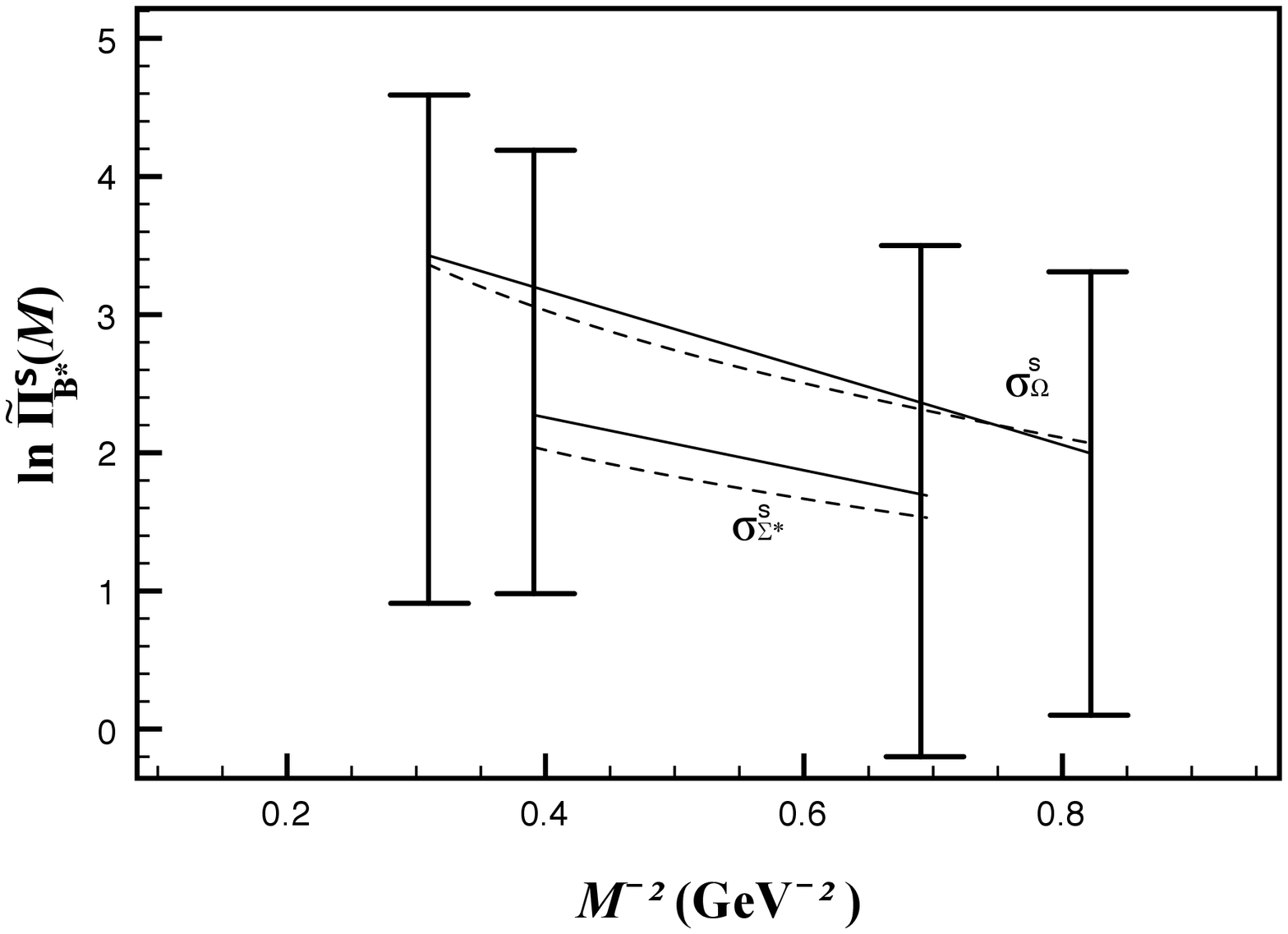}  \label{log_decuplet_s}
\end{figure}

To demonstrate how well the sum rules and the fitting work, we first arrange the sum rules in the subtracted form 
{\allowdisplaybreaks
\begin{subequations}\label{subforms}
	\begin{align}
			\tilde{\Pi}_{{\cal B}^{(\ast)}} &\equiv \tilde{\lambda}_{{\cal B}^{(\ast)}}^2 m_{{\cal B}^{(\ast)}}\frac{\sigma_{{\cal B}^{(\ast)}}}{\hat{m}} e^{-m_{{\cal B}^{(\ast)}}^2/M^2},\\
			\tilde{\Pi}_{{\cal B}^{(\ast)}}^{s} &\equiv \tilde{\lambda}_{{\cal B}^{(\ast)}}^2 m_{B^{(\ast)}}\frac{\sigma^s_{{\cal B}^{(\ast)}}}{m_s} e^{-m_{{\cal B}^{(\ast)}}^2/M^2},
	\end{align}%
\end{subequations}
}%
where $\tilde{\Pi}_{{\cal B}^{(\ast)}}$ and $\tilde{\Pi}_{{\cal B}^{(\ast)}}^{s}$ represent the OPE-minus-excited state and minus-continuum-change contributions, and then we plot the logarithms of both sides. As the RHS appears as a straight line with this form, the linearity of the LHS gives an indication of the OPE convergence and the quality of the continuum model. This procedure is equivalent to searching for a plateau region as a function of the Borel mass as in the `traditional' analysis of the QCDSR. Figs.~\ref{log_octet}-\ref{log_decuplet_s} show the logarithms of the subtracted forms in \eqref{subforms} as a function of inverse Borel-mass squared. Almost linear behavior of the subtracted forms in these figures implies that the valid Borel regions selected according to the criterion above match the plateau regions.

\begin{figure*}
	[th] 
	\caption{(color online) Scatter plots showing the correlations between $\sigma_\Sigma$, $\sigma_{\Xi^\ast}$, and $\chi$, $\chi_G$, $\kappa$. The data are normalized with the mean values of the sigma terms (the normalized value is represented by $\tilde{\sigma}_{B^{(\ast)}}$). The shaded regions represent the extracted values of the sigma terms with their errors. The data with the error bars (in blue) are shown for reference and give the value of the sigma terms when $\chi$, $\chi_G$ or $\kappa$ deviate by 3 standard deviations.}
	\includegraphics[scale=0.40]{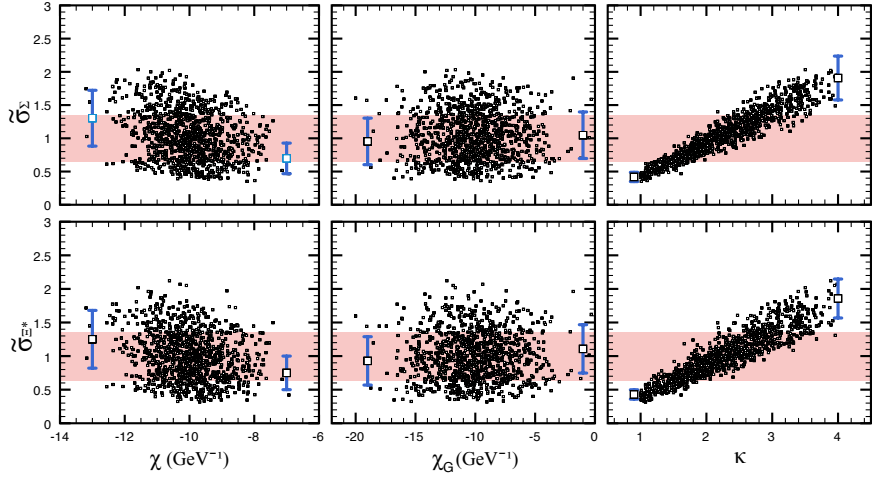}  \label{scatter}
\end{figure*}

\begin{figure*}
	[th] 
	\caption{(color online) Same as Fig.~\ref{scatter} but for the correlations between $\sigma^s_\Xi$, $\sigma^s_{\Xi^\ast}$, and $\chi^s$, $\chi^s_G$, $\kappa$.}
	\includegraphics[scale=0.40]{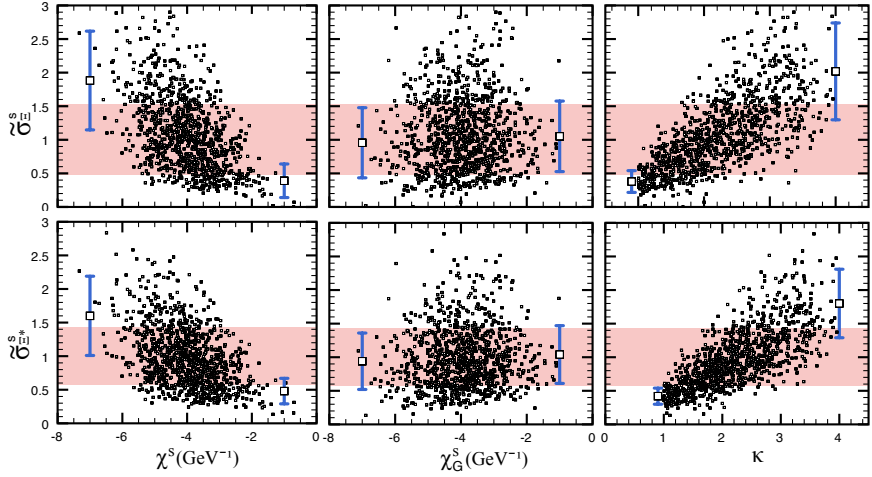} \label{scatter_s}
\end{figure*}
\begin{figure*}
	[th]
	\caption{(color online) Same as Fig.~\ref{scatter} but for the correlations between the ratios $\sigma_\Sigma/\sigma_\Lambda$, $\sigma_{\Xi^\ast}/\sigma_{\Sigma^\ast}$, and $\chi$, $\chi_G$, $\kappa$.}
	\includegraphics[scale=0.40]{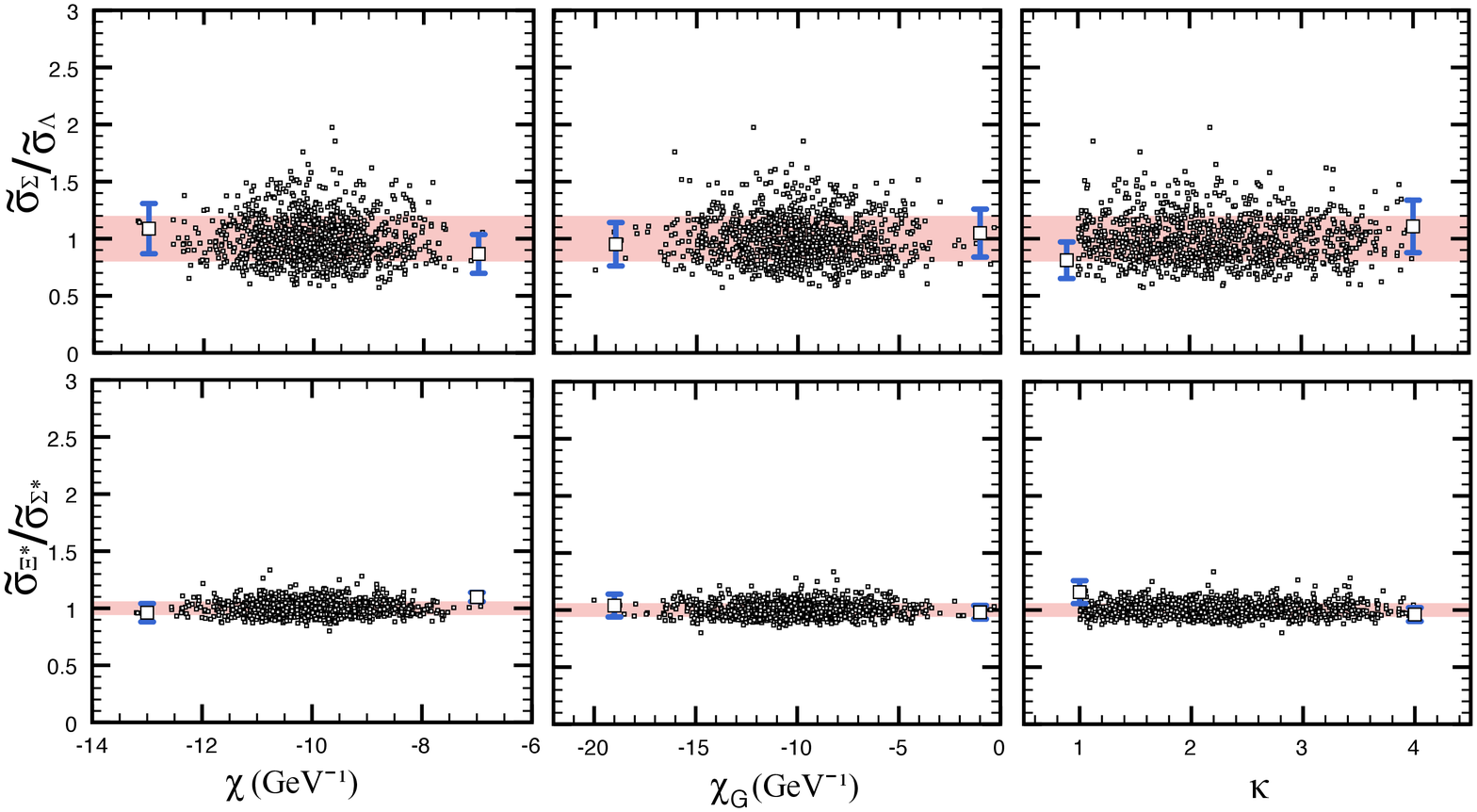} \label{scatter_rat}
\end{figure*}

\begin{figure*}
	[th] 
	\caption{(color online) Same as Fig.~\ref{scatter} but for the correlations between the ratios $\sigma^s_\Xi/\sigma^s_\Lambda$, $\sigma^s_{\Xi^\ast}/\sigma_{\Sigma^\ast}$, and $\chi^s$, $\chi^s_G$, $\kappa$.}
	\includegraphics[scale=0.40]{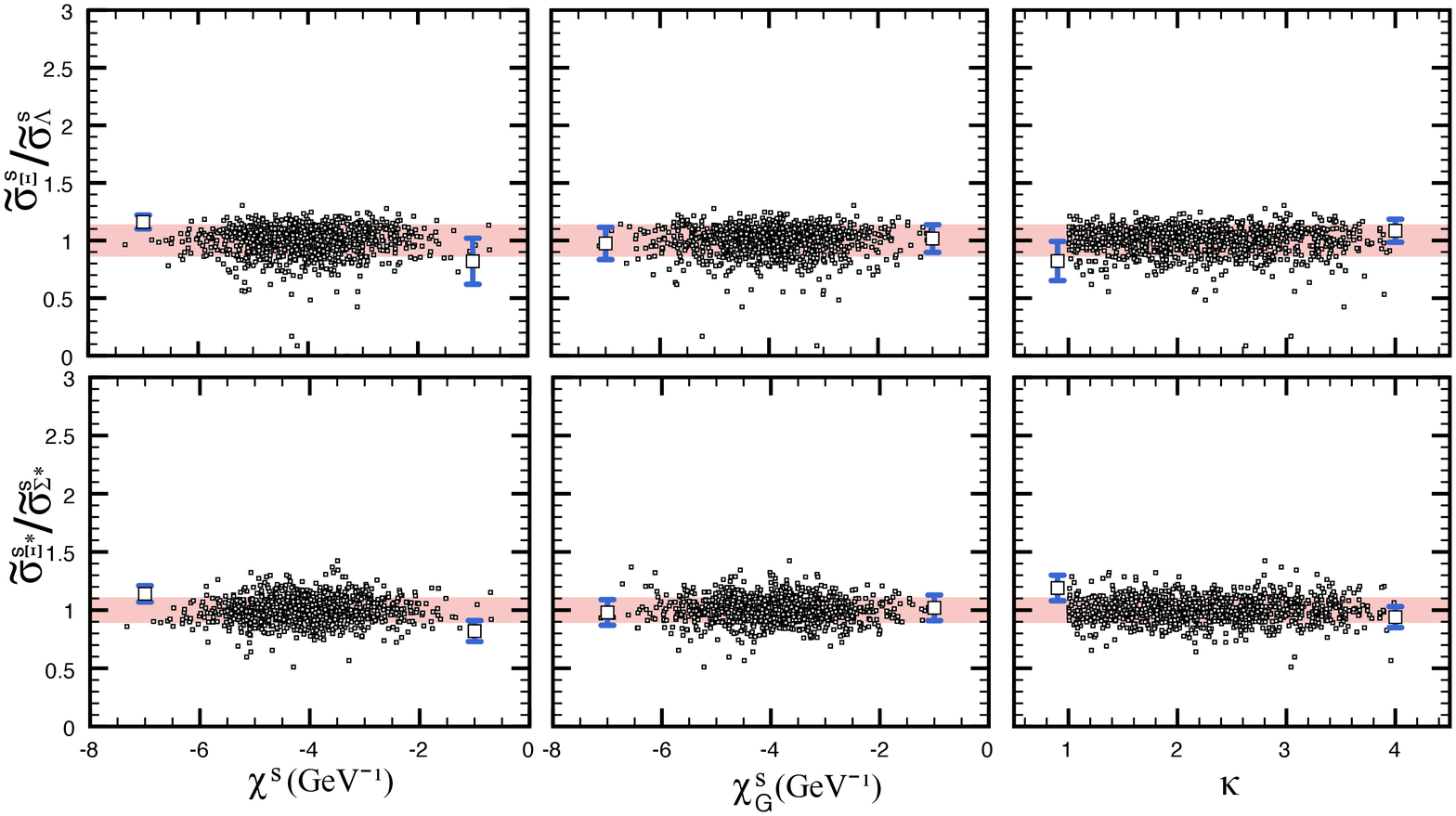} \label{scatter_rat_s}
\end{figure*}

In Table~\ref{ST_table} we present the values of the sigma terms, $\sigma_{B^{(\ast)}}$ and $\sigma^{s}_{B^{(\ast)}}$, and the transition strengths, $\tilde{C}_{B^{(\ast)}}$ and $\tilde{C}^{s}_{B^{(\ast)}}$, for each resonance as obtained from the fits of \eqref{sigsigma}-\eqref{sigomega} with $10^3$ parameter sets. The second column shows the valid Borel regions that are determined according to the criterion explained above and the third column shows the continuum-plus-continuum-change and plus-excited-state contributions at the lower ends of the valid Borel regions. For comparison purposes, we also give the values of the sigma terms as obtained from a chiral model in Ref.~\cite{Barros:2002mt} and from NJL model~\cite{Kunihiro:1990ts,Hatsuda:1994pi} (the latter are obtained by multiplying the quark condensates of the baryons given in Ref.~\cite{Hatsuda:1994pi} with the central values of the quark masses, $\hat{m}=6$~MeV and $m_s=110$~MeV). We quote two errors for the extracted sigma terms: The second is the error due to uncertainty in the $\kappa$ parameter and the first is the sum of errors due to all remaining sources.

The large errors in the final results and the dependence on QCD parameters can be substantially removed by considering the ratios of the sigma terms. We extract these ratios, which are listed in the last column of Table~\ref{ST_table}, by dividing the corresponding values of the two sigma terms for each QCD parameter set and by making a statistical analysis of the final distribution. By comparing the values of the sigma terms with those of the ratios, we find that the ratios can be determined rather accurately due to cancellations in the systematic errors. The errors of $\sim$50\% in the final results are reduced to a level of 10-20\% in most of the cases when the ratios are considered. We demonstrate this fact by studying the correlations between the fit and the QCD parameters via scatter plots. In Fig.~\ref{scatter}, we present the scatter plots showing the correlations between $\sigma_\Sigma$, $\sigma_{\Xi^\ast}$, and $\chi$, $\chi_G$, $\kappa$. In these figures, the data are normalized with the mean values of the sigma terms (the normalized value is represented by $\tilde{\sigma}_{B^{(\ast)}}$) so that the results can be compared on the same scale. The shaded regions represent the extracted values of the sigma terms with their errors. The data with the error bars (in blue) are shown for reference and give the value of the sigma terms when $\chi$, $\chi_G$ or $\kappa$ are changed by 3 standard deviations. In Fig.~\ref{scatter_s}, similar scatter plots are given for the correlations between $\sigma^s_\Xi$, $\sigma^s_{\Xi^\ast}$, and $\chi^s$, $\chi^s_G$, $\kappa$. The Monte Carlo analysis has the advantage that it covers wide ranges of parameter values. We observe no correlation with $\chi_G$ and $\chi^s_G$, which implies that the results are almost independent of $\chi_G$ and $\chi^s_G$. A slight positive correlation is observed with the absolute value of $\chi$, while the correlations with $\chi^s$ are somewhat stronger. The scatter plots showing the correlations between the sigma terms and the $\kappa$ parameter imply that the sum rules have the strongest dependence on $\kappa$. This dependence is also the main source of error in the final results of the sigma terms.

We now turn to the ratios of the sigma terms. In Figs.~\ref{scatter_rat} and~\ref{scatter_rat_s} similar scatter plots are shown for the correlations between the ratios $\sigma_\Sigma/\sigma_\Lambda$, $\sigma_{\Xi^\ast}/\sigma_{\Sigma^\ast}$, $\sigma^s_\Xi/\sigma_\Lambda$, $\sigma^s_{\Xi^\ast}/\sigma_{\Sigma^\ast}$ and the parameters $\chi$, $\chi^s$, $\chi_G$, $\chi^s_G$, $\kappa$. It is impressive to observe that the errors are reduced to a great extent when the ratios are considered and only a slight dependence remains on the QCD parameters. Even the strongest dependence on $\kappa$ parameter is removed. This behavior is common to all sigma terms, which suggests that more accurate and reliable results can be obtained by considering the ratios.

\section{Conclusions and discussion}~\label{secCONC}
We have derived the QCDSR for the scalar quark condensates of the octet and the decuplet baryons. This, together with the definitions in Eqs.~(\ref{sff}) and~\eqref{sffs}, leads to a determination of meson-baryon sigma terms. We have applied a Monte Carlo-based analysis of the sum rules, where we have generated randomly-selected parameter sets from the uncertainties in the QCD input parameters and have made $10^3$ fits in order to determine how the initial errors propagate to the final fit parameters. To determine the valid Borel windows, we have applied two criteria which take account of the pole dominance and the OPE convergence.

The large errors in the fit parameters originate from several sources. The factorization violation parameter, $\kappa$, and the susceptibilities are the main sources of uncertainty. We have determined the susceptibility $\chi$ of the light quark sector in a model independent way by using the chiral perturbation theory results. The susceptibility $\chi^s$ can be related to $\chi$ by using the three-flavor NJL model. It is reasonable to assume $\chi_G^s \equiv \chi^s$ and $\chi_G \equiv \chi$ (with larger uncertainties). Our analysis shows that the sigma terms have no considerable dependence on $\chi_G$ and $\chi^s_G$: The results are consistent with the current ones even for the vanishing values of these susceptibilities. We consider anomalous quark contents of the octet and the decuplet baryons by allowing small but non-zero values for $\tilde{\chi}$, $\tilde{\chi}^s$, $\tilde{\chi}_G$ and $\tilde{\chi}_G^s$. Note that such an OZI-rule--violating case is considered as a trial complementary to our analysis. In the limit of vanishing non-diagonal susceptibilities, the sigma terms $\sigma_\Omega$, $\sigma^s_N$ and $\sigma^s_\Delta$ vanish as well (in consistency with the OZI rule) and all other sigma terms and their ratios are mostly unaffected.

The overlap amplitudes as determined from the baryon-mass sum rules, which we have used to normalize the sum rules for the sigma terms, introduce considerable uncertainties and can only be improved by a better accuracy of the mass sum rules as a result of reducing the errors in the input QCD parameters. Lastly, the transitions to higher-order states and the unknown change of the continuum thresholds with the external field are another sources of uncertainty. We have observed that although a generous range is allowed for the second, it has relatively small impact on the final results and can be safely neglected. However, the transitions to higher-order states, which are not properly suppressed after the Borel transformation, should be included on the phenomenological side as they give large contributions for most of the cases. 

It is impressive to see that the large systematic errors cancel when we consider the ratios of the sigma terms. We would like to stress the ratio values given in Table~\ref{ST_table} as our main results, which can be rather accurately determined and have a minimal dependence on the QCD parameters. We have demonstrated this fact by studying the correlations between the input parameters and the sigma terms together with their ratios. We have found that the ratios depend very weakly on the $\kappa$ parameter as well as on the susceptibilities (within a wide range of 3 standard deviations), as a result of which the error bars shrink. Our results predict the orderings $\sigma_N \geq \sigma_\Sigma > \sigma_\Lambda > \sigma_\Xi$ and $\sigma_\Delta \sim \sigma_{\Sigma^\ast} > \sigma_{\Xi^\ast}> \sigma_\Omega$ for the pion-baryon sigma terms, independently of the values of the susceptibilities and $\kappa$. As for the strange-quark mass contributions to the octet and decuplet baryons, we find $\sigma^s_\Xi > \sigma^s_\Lambda > \sigma^s_N \geq \sigma^s_\Sigma$ and $\sigma^s_\Omega \geq \sigma^s_{\Sigma^\ast} \sim \sigma^s_{\Xi^\ast} > \sigma^s_\Delta$.

The sigma term gives a measure of the contribution of explicit chiral-symmetry breaking in the baryon masses. The QCD Hamiltonian consists of the chiral-invariant terms containing the gauge couplings of gluons and the chiral--non-invariant quark-mass term. Suppose that the chiral
non-invariant term is weak and therefore treated perturbatively. Then the sigma term is nothing but the contribution of the quark-mass term to the baryon mass. We find that, among the octet and decuplet hyperons, the chiral-symmetry breaking gives the largest contributions to $\Sigma$ and $\Sigma^\ast$ baryons. A non-trivial outcome is that $\sigma_\Xi$ and $\sigma^s_\Sigma$ depend mainly on the non-diagonal responses of the quark condensates (besides some small contributions from direct couplings). These sigma terms are consistent with zero in the limit of vanishing non-diagonal susceptibilities. We also find that $\sigma_\Lambda$ is considerably smaller than $\sigma_\Sigma$, which indicates that the quark mass term contributes to $\Lambda$-$\Sigma$ mass splitting. One caveat in the treatment of the octet baryons is that instead of the optimum choices for the interpolating fields, which are known to perform rather successfully in the mass determination~\cite{Ioffe:1981kw}, it is possible to adopt a generalized definition in terms of arbitrary mixings between two different local operators. Such an extension of our analysis with more general interpolating fields is desirable, but has the difficulty of treating one extra parameter. Note that in the case of the decuplet baryons we have a unique local operator.

Finally, we would like to comment on the anomalous quark content of the baryons. Our sum rules show that a non-vanishing response of $\langle \overline{q}q \rangle$ to external $S_s(\overline{s}s)$ field implies OZI-rule violating $\overline{s}s$ content for the nucleon and the Delta. Similarly, a non-diagonal response of $\langle \overline{s}s \rangle$ to external $S(\overline{u}u+\overline{d}d)$ field leads to a small but non-negligible $\overline{u}u$ content for the Omega. Therefore, the interesting question of baryon anomalous quark contents boils down to a determination of the susceptibilities $\tilde{\chi}$, $\tilde{\chi}^s$, $\tilde{\chi}_G$, and $\tilde{\chi}^s_G$. In the range of the values considered for $\tilde{\chi}$ and $\tilde{\chi}_G$, the value we obtain for the strangeness content of the nucleon as $\sigma^s_N=161\pm 66$~MeV is larger than that from the NJL model but compares favorably to expectations based on chiral perturbation theory ($\simeq 130$~MeV according to Ref.~\cite{Gasser:1990ce}), and on lattice QCD ($183 \pm 8$~MeV according to Ref.~\cite{Dong:1995ec}). This corresponds to a nucleon strangeness fraction of $y=0.34\pm 0.06$ as defined in Eq.~\eqref{stfr}, in agreement with that from lattice QCD as $y \simeq 0.36$~\cite{Dong:1995ec}. The ratio of the OZI-rule violating contributions to the nucleon and the Delta is of special interest:
\begin{equation}\label{ratnd}
	 \frac{\sigma^s_N}{\sigma^s_\Delta}=0.99\pm 0.19,	
\end{equation} 
which predicts an equivalent strangeness content for the nucleon and the Delta. 
\acknowledgments
This work has been supported in part by the Japan Society for the Promotion of Science under contract number P06327 and in part by KAKENHI, 17070002 (Priority area) and 19540275.

\begin{widetext}
\appendix
\section{The QCD sum rules for the meson-baryon sigma terms}\label{app1}
In this Appendix, we give the sum rules for the meson-baryon sigma terms, which are obtained by matching the OPE sides with the hadronic sides and applying the Borel transformation:
{\allowdisplaybreaks

	\begin{align}
		\begin{split}\label{sigsigma}
			&\sigma_\Sigma:\\
			&-\frac{4}{3}\kappa \,\chi\,a_q^2\,M^2\,L^{4/9} -\frac{m_0^2}{3}\,a_q\,M^2\,L^{-14/27} +\frac{m_0^2}{6} (\chi_G+\chi) \,a_q^2 \,L^{-2/27}\\
			&\qquad +m_s\tilde{\chi}\,f\,a_q\,M^4\, E^\Sigma_0\,L^{-4/9} +\frac{m_s}{6}\tilde{\chi}_G\,m_0^2 \,a_q\,f\,M^2\,L^{-26/27}\\
			&\quad =\left[\tilde{\lambda}_{\Sigma}^2 m_{\Sigma} \frac{\sigma_{\Sigma}}{\hat{m}}+\,\tilde{C}_{\Sigma}\,M^2 +\, \left(\frac{w_{\Sigma}^4}{2}- 2 m_s\,f\, a_q \right) \delta w_{\Sigma}^2\,M^2 \,L^{-4/9} e^{(m_{\Sigma}^2-w_{\Sigma}^2)/M^2} \right]\,e^{-m_{\Sigma}^2/M^2},
		\end{split}\\
		\begin{split}\label{siglambda}
			\raisetag{80pt}
			&\sigma_\Lambda:\\
			&\frac{4}{3}\,a_q\,(f-2)\,M^4 \,E^\Lambda_0- \frac{4}{9}\kappa\,\chi \,a_q^2(2f-1)\,M^2\,L^{4/9}- \frac{8}{9}\kappa\,\tilde{\chi}\,f \,a_q^2\,M^2\,L^{4/9} \\
			&\qquad +\frac{m_0^2}{3}\, f\, a_q\,M^2\,L^{-14/27} +\frac{m_0^2}{18} (\chi +\chi_G)\,a_q^2(2f-1)\,L^{-2/27}+\frac{m_0^2}{9} (\tilde{\chi} +\tilde{\chi}_G)\,f\,a_q^2\,L^{-2/27}  \\
			&\qquad  + \frac{4 m_s}{3}\,M^6\,E^\Lambda_1\,L^{-8/9} -\frac{4 m_s}{3}\chi\,a_q\,M^4\, E^\Lambda_0\,L^{-4/9} +m_s\tilde{\chi}\,f\,a_q\,M^4\, E^\Lambda_0\,L^{-4/9}\\
			&\qquad -\frac{m_s}{3}\chi_G\,m_0^2\,a_q\,M^2\,L^{-26/27} -\frac{m_s}{6}\tilde{\chi}_G\,m_0^2\,a_q \,f\,M^2\,L^{-26/27} +\frac{4m_s}{9}(2f-1)\kappa\,a_q^2\\
			&\quad =\left[\tilde{\lambda}_{\Lambda}^2 m_{\Lambda} \frac{\sigma_{\Lambda}}{\hat{m}}+\,\tilde{C}_{\Lambda}\,M^2 +\, \left(\frac{w_{\Lambda}^4}{2}- \frac{3m_s}{4} (3f-4)\, a_q \right) \delta w_{\Lambda}^2\,M^2 \,L^{-4/9} \right.\\
			&\qquad \left. \times e^{(m_{\Lambda}^2-w_{\Lambda}^2)/M^2} \right]\,e^{-m_{\Lambda}^2/M^2},
		\end{split}\\
		\begin{split}\label{sigxi}
			&\sigma_\Xi:\\
			&-a_q\,M^4\,E^\Xi_0-\frac{m_0^2}{6}a_q \,M^2\,L^{-14/27}-\frac{4}{3}\kappa\,\tilde{\chi}\,f^2 \,a_q^2\,M^2\,L^{4/9}\\
			&\qquad +\frac{m_0^2}{6} (\tilde{\chi} +\tilde{\chi}_G)\,f^2\,a_q^2\,L^{-2/27} +\frac{m_s}{3}\tilde{\chi}_G\,m_0^2\,a_q\,f\,M^2\,L^{-26/27}\\
			&\quad =\left(\tilde{\lambda}_{\Xi}^2 m_{\Xi} \frac{\sigma_{\Xi}}{\hat{m}}+\,\tilde{C}_{\Xi}\,M^2 +\, \frac{M^2}{2} w_{\Xi}^4\, \delta w_{\Xi}^2 \,L^{-4/9} e^{(m_{\Xi}^2-w_{\Xi}^2)/M^2} \right)\,e^{-m_{\Xi}^2/M^2},
		\end{split}\\
		\begin{split}\label{sigsigmaast}
			&\sigma_{\Sigma^\ast}:\\
			&\frac{2}{3}a_q\,(1+2f)\, M^4\, E^{\Sigma^\ast}_0\, L^{16/27}-\frac{8}{9} \chi\, \kappa\, (1+f)\, a_q^2\, M^2\,L^{28/27}-\frac{8}{9} \tilde{\chi}\, \kappa\, f\, a_q^2\, M^2\, L^{28/27}\\
			&\qquad-\frac{m_0^2}{9} (2+7f)\, a_q\, M^2\,L^{2/27}  +\frac{7}{27} (1+f)\, (\chi+\chi_G)\, m_0^2\, a_q^2\, L^{14/27}\\
			&\qquad +\frac{7}{27} f\, (\tilde{\chi}+\tilde{\chi}_G)\, m_0^2\, a_q^2\, L^{14/27} + \frac{4 m_s}{3} M^6\, E^{\Sigma^\ast}_1\, L^{-8/27} -\frac{4 m_s}{3}  \chi\, a_q\, M^4\,E^{\Sigma^\ast}_0\, L^{4/27} \\
			&\qquad +\frac{m_s}{3}  \tilde{\chi}\,f a_q\, M^4\,E^{\Sigma^\ast}_0\, L^{4/27} +\frac{7 m_s}{9} \chi_G\, m_0^2\, a_q\, M^2\,L^{-10/27}\\
			&\qquad -\frac{5 m_s}{18} \tilde{\chi}_G\,f\, m_0^2\, a_q\, M^2\,L^{-10/27} - \frac{4 m_s}{9} (1+f)\, \kappa\, a_q^2\, L^{16/27} \\
			&\quad =\left[\tilde{\lambda}_{\Sigma^\ast}^2 m_{\Sigma^\ast} \frac{\sigma_{\Sigma^\ast}}{\hat{m}}+\,\tilde{C}_{\Sigma^\ast}\,M^2 +\, \left(\frac{w_{\Sigma^\ast}^4}{5}-\frac{2m_s}{3} (4-f)\, a_q \right) \delta w_{\Sigma^\ast}^2\, M^2 \,L^{4/27} \right. \\
			&\qquad \left.\times e^{(m_{\Sigma^\ast}^2-w_{\Sigma^\ast}^2)/M^2} \right] e^{-m_{\Sigma^\ast}^2/M^2},
		\end{split}\\
		\begin{split}\label{sigxiast}
			&\sigma_{\Xi^\ast}:\\
			&\frac{a_q}{3} M^4\,(4f-1)\, E^{\Xi^\ast}_0\, L^{16/27}-\frac{8}{9} \chi\, \kappa\, f\, a_q^2\, M^2\, L^{28/27}- \frac{8}{9} \tilde{\chi}\, \kappa\, f (1+f)\, a_q^2\, M^2\,L^{28/27} \\
			&\qquad +\frac{m_0^2}{18} (5-14f)\, a_q\, M^2\,L^{2/27} +\frac{7}{27} f\, (\chi+\chi_G)\, m_0^2\, a_q^2\, L^{14/27}  \\
			&\qquad +\frac{7}{27} f(1+f)\, (\tilde{\chi}+\tilde{\chi}_G)\, m_0^2\, a_q^2\, L^{14/27}+ \frac{4 m_s}{3} M^6\, E^{\Xi^\ast}_1\, L^{-8/27}   \\
			&\qquad -\frac{4 m_s}{3}  \chi\, a_q\, M^4\,E^{\Xi^\ast}_0\, L^{4/27} -\frac{2 m_s}{3} f \tilde{\chi}\, a_q\, M^4\,E^{\Xi^\ast}_0\, L^{4/27} +\frac{7 m_s}{9} \chi_G\, m_0^2\, a_q\,M^2\, L^{-10/27} \\
			&\qquad  +\frac{2 m_s}{9} \tilde{\chi}_G\, m_0^2\,f a_q\, M^2\,L^{-10/27} - \frac{4 m_s}{9} (1+f)\, \kappa\, a_q^2\,L^{16/27} \\
			&\quad =\left[\tilde{\lambda}_{\Xi^\ast}^2 m_{\Xi^\ast} \frac{\sigma_{\Xi^\ast}}{\hat{m}}+\,\tilde{C}_{\Xi^\ast}\,M^2 +\, \left(\frac{w_{\Xi^\ast}^4}{5}-\frac{4m_s}{3} (f+2)\, a_q \right) \delta w_{\Xi^\ast}^2\, M^2 \,L^{4/27} \right. \\
			&\qquad \left.\times e^{(m_{\Xi^\ast}^2-w_{\Xi^\ast}^2)/M^2} \right] e^{-m_{\Xi^\ast}^2/M^2},
		\end{split}\\
		\begin{split}\label{sigomegaud}
			&\sigma_{\Omega}:\\
			&-\frac{8}{3} \tilde{\chi}\, \kappa\, f^2\, a_q^2\, M^2\,L^{28/27} +\frac{7}{9} f^2\, (\tilde{\chi}+\tilde{\chi}_G)\, m_0^2\, a_q^2\, L^{14/27}\\
			&\qquad -3 m_s\, f \tilde{\chi}\, a_q\, M^4\,E^\Omega_0\, L^{4/27} +\frac{3 m_s}{2} \tilde{\chi}_G\, m_0^2\,f a_q\, M^2\,L^{-10/27} \\
			&\quad =\left[\tilde{\lambda}_{\Omega}^2 m_{\Omega} \frac{\sigma_\Omega}{\hat{m}}+\,\tilde{C}_\Omega\,M^2 +\, \left(\frac{w_\Omega^4}{5}- 6 m_s\,f\, a_q \right) \delta (w_\Omega)^2\, M^2 \,L^{4/27} \right. \\
			&\qquad \left.\times e^{(m_{\Omega}^2-(w_\Omega)^2)/M^2} \right] e^{-m_\Omega^2/M^2},
		\end{split}	\\
		\begin{split}\label{sigN}
			&\sigma^s_N:\\
			&-\frac{4}{3}\kappa \,\tilde{\chi}^s\,a_q^2\,M^2\,L^{4/9} +(\tilde{\chi}^s+\tilde{\chi}^s_G) \frac{m_0^2}{6} \,a_q^2 \,L^{-2/27}\\
			&\quad =\left[\tilde{\lambda}_N^2 m_N \frac{\sigma^s_N}{m_s}+\,\tilde{C}^s_N\,M^2 +\, \frac{(w^s_N)^4}{2} \delta (w^s_N)^2\,M^2 \,L^{-4/9} e^{(m_N^2-(w^s_N)^2)/M^2} \right]\,e^{-m_N^2/M^2},
		\end{split}\\
		\begin{split}\label{sigsigmas}
			&\sigma^s_\Sigma:\\
			&-a_q\,f\,M^4\,\tilde{E}^\Sigma_0 -\frac{m_0^2}{6}\,f\,a_q\,M^2\,L^{-14/27}-\frac{4}{3}\kappa \,\tilde{\chi}^s\,a_q^2\,M^2\,L^{4/9} +(\tilde{\chi}^s_G+\tilde{\chi}^s) \frac{m_0^2}{6} \,a_q^2 \,L^{-2/27}\\
			&\qquad -4 m_s\,M^6\,\tilde{E}^\Sigma_1\,L^{-8/9}+m_s\chi^s \,f\,a_q\,M^4\, \tilde{E}^\Sigma_0\,L^{-4/9} +\frac{m_s}{6}\chi^s_G\,m_0^2 \,a_q\,f\,M^2\,L^{-26/27}\\
			&\qquad -\frac{4m_s}{3}\kappa\,a_q^2\\
			&\quad =\left[\tilde{\lambda}_{\Sigma}^2 m_{\Sigma} \frac{\sigma^s_{\Sigma}}{m_s}+\,\tilde{C}^s_{\Sigma}\,M^2 +\, \left(\frac{(w^s_{\Sigma})^4}{2}- 2 m_s\,f\, a_q \right) \delta (w^s_{\Sigma})^2\,M^2 \,L^{-4/9} \right.\\
			&\qquad \left. \times e^{(m_{\Sigma}^2-(w^s_{\Sigma})^2)/M^2} \right]\,e^{-m_{\Sigma}^2/M^2},
		\end{split}\\
	\begin{split}\label{siglambdas}
			&\sigma^s_\Lambda:\\
			&\frac{a_q}{3}(4-3f)\,M^4 \,\tilde{E}^\Lambda_0 - \frac{8}{9}\kappa\,\chi_s\,f \,a_q^2\,M^2\,L^{4/9}- \frac{4}{9}\kappa\,\tilde{\chi}^s \,a_q^2(2f-1)\,M^2\,L^{4/9} \\
			&\qquad+\frac{m_0^2}{6}\, (f-2)\, a_q\,M^2\,L^{-14/27}+\frac{m_0^2}{9} (\chi^s +\chi^s_G)\,f\,a_q^2\,L^{-2/27}\\
			&\qquad+\frac{m_0^2}{18} (\tilde{\chi}^s +\tilde{\chi}^s_G)\,a_q^2(2f-1)\,L^{-2/27} + 4 m_s\,M^6\,\tilde{E}^\Lambda_1\,L^{-8/9} +m_s\chi_s\,f\,a_q\,M^4\, \tilde{E}^\Lambda_0\,L^{-4/9}\\
			&\qquad-\frac{4 m_s}{3}\tilde{\chi}^s\,a_q\,M^4\, E^\Lambda_0\,L^{-4/9}-\frac{m_s}{6}\chi^s_G\,m_0^2\,a_q\,f\,M^2\,L^{-26/27} -\frac{m_s}{3}\tilde{\chi}^s_G\,m_0^2\,a_q\,M^2\,L^{-26/27} \\
			&\qquad+\frac{4m_s}{9}\,f\,\kappa\,a_q^2\\
			&\quad =\left[\tilde{\lambda}_{\Lambda}^2 m_{\Lambda} \frac{\sigma^s_{\Lambda}}{m_s}+\,\tilde{C}^s_{\Lambda}\,M^2 +\, \left(\frac{(w^s_{\Lambda})^4}{2}- \frac{3m_s}{4} (3f-4)\, a_q \right)  \delta (w^s_{\Lambda})^2\,M^2 \,L^{-4/9} \right.\\
			&\qquad \left. \times e^{(m_{\Lambda}^2-(w^s_{\Lambda})^2)/M^2} \right]\,e^{-m_{\Lambda}^2/M^2},
		\end{split}\\
		\begin{split}\label{sigxis}
			&\sigma^s_\Xi:\\
			&-\frac{m_0^2}{3}a_q\,f\,M^2\,L^{-14/27}+\frac{m_0^2}{6} (\chi^s +\chi^s_G)\,f^2\,a_q^2\,L^{-2/27}-\frac{4}{3}\kappa\,\chi^s\,f^2 \,a_q^2\,M^2\,L^{4/9}\\
			&\qquad - 6 m_s\,M^6\,\tilde{E}^\Xi_1\,L^{-8/9} +\frac{m_s}{3}\chi^s_G\,m_0^2\,a_q\,f\,M^2\,L^{-26/27} +\frac{4m_s}{3}\,f^2\,\kappa\,a_q^2\\
			&\quad =\left(\tilde{\lambda}_{\Xi}^2 m_{\Xi} \frac{\sigma^s_{\Xi}}{m_s}+\,\tilde{C}^s_{\Xi}\,M^2 +\, \frac{M^2}{2} (w^s_{\Xi})^4\, \delta (w^s_{\Xi})^2 \,L^{-4/9} e^{(m_{\Xi}^2-(w^s_{\Xi})^2)/M^2} \right)\,e^{-m_{\Xi}^2/M^2},
		\end{split}	\\
		\begin{split}\label{sigDeltas}
			&\sigma^s_{\Delta}:\\
			&-\frac{8}{3} \tilde{\chi}^s\, \kappa\, a_q^2\, M^2\,L^{28/27} +\frac{7}{9} (\tilde{\chi}^s+\tilde{\chi}^s_G)\, m_0^2\, a_q^2\, L^{14/27}\\
			&\quad =\left[\tilde{\lambda}_{\Delta}^2 m_{\Delta} \frac{\sigma^s_\Delta}{m_s}+\,\tilde{C}^s_\Delta\,M^2 +\, \frac{(w_\Delta^s)^4}{5} \delta (w^s_\Delta)^2\, M^2 \,L^{4/27} e^{(m_{\Delta}^2-(w^s_\Delta)^2)/M^2} \right] e^{-m_\Delta^2/M^2},
		\end{split}\\
		\begin{split}\label{sigsigmaasts}
			&\sigma^s_{\Sigma^\ast}:\\
			&\frac{a_q}{3} M^4\,(4-f)\, \tilde{E}^{\Sigma^\ast}_0\, L^{16/27}-\frac{8}{9} \chi^s\, \kappa\, f\, a_q^2\, M^2\, L^{28/27}-\frac{8}{9} \tilde{\chi}^s\, \kappa\, (1+f)\, a_q^2\, M^2\,L^{28/27}\\
			&\qquad +\frac{m_0^2}{18} (5f-14)\, a_q\, M^2\,L^{2/27} +\frac{7}{27} f\, (\chi^s+\chi^s_G)\, m_0^2\, a_q^2\, L^{14/27}\\
			&\qquad +\frac{7}{27} (1+f)\, (\tilde{\chi}^s+\tilde{\chi}^s_G)\, m_0^2\, a_q^2\, L^{14/27} - \frac{3 m_s}{2} M^6\, \tilde{E}^{\Sigma^\ast}_1\, L^{-8/27} \\
			&\qquad +\frac{m_s}{3} \chi^s\,f a_q\, M^4\,\tilde{E}^{\Sigma^\ast}_0\, L^{4/27}-\frac{4 m_s}{3}  \tilde{\chi}^s\, a_q\, M^4\,\tilde{E}^{\Sigma^\ast}_0\, L^{4/27} -\frac{5 m_s}{18} \chi^s_G m_0^2\,f\, a_q\, M^2\,L^{-10/27} \\
			&\qquad +\frac{7 m_s}{9} \tilde{\chi}^s_G\, m_0^2\, a_q\, M^2\,L^{-10/27}+\frac{4 m_s}{9} (f-2)\, \kappa\, a_q^2\,L^{16/27} \\
			&\quad =\left[\tilde{\lambda}_{\Sigma^\ast}^2 m_{\Sigma^\ast} \frac{\sigma_{\Sigma^\ast}}{m_s}+\,\tilde{C}^s_{\Sigma^\ast}\,M^2 +\, \left(\frac{(w^s_{\Sigma^\ast})^4}{5}-\frac{2m_s}{3} (4-f)\, a_q \right) \delta (w_{\Sigma^\ast})^2\, M^2 \,L^{4/27} \right. \\
			&\qquad \left.\times e^{(m_{\Sigma^\ast}^2- (w^s_{\Sigma^\ast})^2)/M^2} \right] e^{-m_{\Sigma^\ast}^2/M^2}, \raisetag{80pt}
		\end{split}\\
		\begin{split}\label{sigxiasts}
			&\sigma^s_{\Xi^\ast}:\\
			&\frac{2}{3}a_q\,(2+f)\, M^4\, \tilde{E}^{\Xi^\ast}_0\, L^{16/27}-\frac{8}{9} \chi^s\, \kappa\, f (1+f)\, a_q^2\, M^2\,L^{28/27} -\frac{8}{9} \tilde{\chi}^s\, \kappa\, f\, a_q^2\, M^2\,L^{28/27} \\
			&\qquad -\frac{m_0^2}{9} (7+2f)\, a_q\, M^2\,L^{2/27} +\frac{7}{27} f(1+f)\, (\chi^s+\chi^s_G)\, m_0^2\, a_q^2\, L^{14/27}  \\
			&\qquad +\frac{7}{27} f\, (\tilde{\chi}^s+\tilde{\chi}^s_G)\, m_0^2\, a_q^2\, L^{14/27}- \frac{5 m_s}{3} M^6\, \tilde{E}^{\Xi^\ast}_1\, L^{-8/27}-\frac{2 m_s}{3} f \chi^s\, a_q\, M^4\,\tilde{E}^{\Xi^\ast}_0\, L^{4/27}  \\
			&\qquad -\frac{4 m_s}{3} \tilde{\chi}^s\, a_q\, M^4\,\tilde{E}^{\Xi^\ast}_0\, L^{4/27}+\frac{2 m_s}{9} \chi^s_G\, m_0^2\,f a_q\, M^2\,L^{-10/27} \\
			&\qquad +\frac{7 m_s}{9} \tilde{\chi}^s_G\, m_0^2\, a_q\,M^2\, L^{-10/27}- \frac{4 m_s}{9} f(5-f)\, \kappa\, a_q^2\, L^{16/27} \\
			&\quad =\left[\tilde{\lambda}_{\Xi^\ast}^2 m_{\Xi^\ast} \frac{\sigma_{\Xi^\ast}}{m_s}+\,\tilde{C}^s_{\Xi^\ast}\,M^2 +\, \left(\frac{(w_{\Xi^\ast}^s)^4}{5}-\frac{4m_s}{3} (f+2)\, a_q \right) \delta (w^s_{\Xi^\ast})^2\, M^2 \,L^{4/27} \right. \\
			&\qquad \left.\times e^{(m_{\Xi^\ast}^2-(w^s_{\Xi^\ast})^2)/M^2} \right] e^{-m_{\Xi^\ast}^2/M^2},
		\end{split}	\\
		\begin{split}\label{sigomega}
			&\sigma^s_{\Omega}:\\
			&3 a_q\,f\, M^4\, \tilde{E}^\Omega_0\, L^{16/27}-\frac{8}{3} \chi^s\, \kappa\, f^2\, a_q^2\, M^2\,L^{28/27} -\frac{3}{2} f\, m_0^2\,a_q\, M^2\,L^{2/27}\\
			&\qquad +\frac{7}{9} f^2\, (\chi^s+\chi^s_G)\, m_0^2\, a_q^2\, L^{14/27} - \frac{m_s}{2} M^6\, \tilde{E}^\Omega_1\, L^{-8/27} - 3 m_s\, f \chi^s\, a_q\, M^4\,\tilde{E}^\Omega_0\, L^{4/27}\\
			&\qquad+\frac{3 m_s}{2} \chi_G^s\, m_0^2\,f a_q\, M^2\,L^{-10/27} - 4 m_s\, f^2\, \kappa\, a_q^2\, L^{16/27} \\
			&\quad =\left[\tilde{\lambda}_{\Omega}^2 m_{\Omega} \frac{\sigma^s_\Omega}{m_s}+\,\tilde{C}^s_\Omega\,M^2 +\, \left(\frac{(w_\Omega^s)^4}{5}- 6 m_s\,f\, a_q \right) \delta (w^s_\Omega)^2\, M^2 \,L^{4/27} \right. \\
			&\qquad \left.\times e^{(m_{\Omega}^2-(w^s_\Omega)^2)/M^2} \right] e^{-m_\Omega^2/M^2}.
		\end{split}
	\end{align}%
}%
Here $M$ is the Borel mass and we have defined $\tilde{\lambda}_{\cal B^{(\ast)}}^2=32 \pi^4 \lambda_{\cal B^{(\ast)}}^2$ and $\tilde{C}^{(s)}_{\cal B^{(\ast)}}=16 \pi^4 C^{(s)}_{\cal B^{(\ast)}}$. 

\section{The QCD sum rules for the baryon masses}\label{app2}
We use the following chiral-odd mass sum rules for normalization of the sigma-term sum rules~\cite{Belyaev:1982sa,Hwang:1994vp,Lee:1997ix}: 
{\allowdisplaybreaks
\begin{align}
	\begin{split}\label{Nmass}
		&m_N:\\
		&a_q\, E^N_1\,M^4-\frac{1}{18}a_q\,b =\frac{\tilde{\lambda}_N^2}{2}m_N\,e^{-m_N^2/M^2},
		\end{split}\\
	\begin{split}\label{LambdamassSR}
		&m_\Lambda:\\
		&\frac{M^4}{3} (4-f)a_q\, E^\Lambda_1- \frac{b}{54}(4-f)a_q-\frac{m_s}{3} M^6\,E^\Lambda_2\,L^{-8/9}+ \frac{m_s}{24}b\,M^2\,E^\Lambda_0\,L^{-8/9}\\
		&\quad+\frac{4\,m_s}{9}\kappa\,a_q^2 (3-f)  = \frac{\tilde{\lambda}_\Lambda^2}{2}\, m_\Lambda\, e^{-m_\Lambda^2/M^2},
	\end{split}\\
	\begin{split}\label{SigmamassSR}
		&m_\Sigma:\\
		&f\,a_q M^4\,E^\Sigma_1-\frac{f\,a_q\,b}{18}+m_s M^6\,E^\Sigma_2\,L^{-8/9}- \frac{m_s}{8}b\,M^2 E^\Sigma_0\, L^{-8/9}+ \frac{4m_s}{3} \kappa\,a_q^2\\
		& \quad =\frac{\tilde{\lambda}_\Sigma^2}{2}\, m_\Sigma\, e^{-m_\Sigma^2/M^2},
	\end{split}\\
	\begin{split}\label{CascademassSR}
		&m_\Xi:\\
		&\,a_q M^4\,E^\Xi_1-\frac{\,a_q\,b}{18}+ 2 m_s\, \kappa\,a_q^2 =\frac{\tilde{\lambda}_\Xi^2}{2}\, m_\Xi\, e^{-m_\Xi^2/M^2},
	\end{split}\\
	\begin{split}\label{mDelta}
	&m_\Delta:\\
	&\frac{4}{3}a_q\,E^\Delta_1\,L^{16/27}M^4-\frac{2}{3}E_0\,m_0^2\, a_q\, L^{2/27}M^2 -\frac{1}{18} a_q\, b\, L^{16/27}= \,\frac{\tilde{\lambda}_\Delta^2}{2}m_\Delta\,e^{-m_\Delta^2/M^2} ,
	\end{split}\\
	\begin{split}
	&m_{\Sigma^\ast}:\\
	&\frac{4}{9}(f+2)\,a_q\,M^4\,E^{\Sigma^\ast}_1 \,L^{16/27}-\frac{2}{9}(f+2) m_0^2\,a_q\,M^2\, E^{\Sigma^\ast}_0\,L^{2/27}-\frac{a_q\,b}{54}(f+2)\, L^{16/27}\\
	&\qquad +\frac{m_s}{2} M^6 \,E^{\Sigma^\ast}_2\,L^{-8/27} -\frac{m_s\,b}{24} M^2\, E^{\Sigma^\ast}_0\, L^{-8/27}+ \frac{2\,m_s}{3}\kappa\,a_q^2\, L^{16/27}\\
	&\quad =\frac{\tilde{\lambda}_{\Sigma^\ast}^2}{2}\, m_{\Sigma^\ast}\, e^{-m_{\Sigma^\ast}^2/M^2},
	\end{split}\\
	\begin{split}
		&m_{\Xi^\ast}:\\
		&\frac{4}{9}(2f+1) a_q\,M^4\,E^{\Xi^\ast}_1\,L^{16/27} -\frac{2}{9}(2f+1) m_0^2\,a_q\,M^2\, E^{\Xi^\ast}_0\,L^{2/27} -\frac{a_q\,b}{54}(2f+1)\, L^{16/27}\\
		&\quad + m_s M^6 \,E^{\Xi^\ast}_2\,L^{-8/27}-\frac{m_s\,b}{12} M^2\, E^{\Xi^\ast}_0\, L^{-8/27}+ \frac{4\,m_s}{3}\kappa\,a_q^2\, L^{16/27}= \frac{\tilde{\lambda}_{\Xi^\ast}^2}{2}\, m_{\Xi^\ast}\, e^{-m_{\Xi^\ast}^2/M^2},
	\end{split}\\
	\begin{split}
	&m_{\Omega}:\\\label{OmegamassSR}
	&\frac{4}{3}f\, a_q\,M^4\,E^{\Omega}_1\,L^{16/27} -\frac{2}{3}f\, m_0^2\,a_q\,M^2\, E^{\Omega}_0\,L^{2/27} -\frac{a_q\,b}{18} f\, L^{16/27}+ \frac{3m_s}{2} M^6 \,E^{\Omega}_2\,L^{-8/27}\\
	&\quad -\frac{m_s\,b}{8} M^2\, E^{\Omega}_0\, L^{-8/27}+ 2 m_s \,f^2\,\kappa\,a_q^2\, L^{16/27}= \frac{\tilde{\lambda}_{\Omega}^2}{2}\, m_{\Omega}\, e^{-m_{\Omega}^2/M^2}.
	\end{split}
\end{align}%
}%
\end{widetext}


\end{document}